\documentclass[reqno,10pt,centertags]{amsart} 
\usepackage{amsmath,amsthm,amscd,amssymb,latexsym,esint,upref,stmaryrd,enumerate,color,verbatim,yfonts,multicol}
\calclayout
\usepackage{color,float}
\usepackage{hyperref}

\usepackage{tikz}
\usetikzlibrary{calc,decorations.markings}
\usepackage{pgfplots}
\pgfplotsset{compat=1.15}

\usepackage{graphicx}

\newcommand*{\mailto}[1]{\href{mailto:#1}{\nolinkurl{#1}}}
\newcommand{\arxiv}[1]{\href{http://arxiv.org/abs/#1}{arXiv:#1}}

\newcommand{\R}{{\mathbb R}}
\newcommand{\N}{{\mathbb N}}

\newcommand{\C}{{\mathbb C}}

\newcommand{\bbC}{{\mathbb{C}}}

\newcommand{\bbN}{{\mathbb{N}}}

\newcommand{\bbR}{{\mathbb{R}}}

\newcommand{\bbZ}{{\mathbb{Z}}}

\newcommand{\cF}{{\mathcal F}}

\newcommand{\beq}{\begin{align}}
\newcommand{\enq}{\end{align}}

\renewcommand{\a}{\alpha}
\renewcommand{\b}{\beta}
\newcommand{\g}{\gamma}
\renewcommand{\d}{\delta}

\renewcommand{\l}{\lambda}

\newcommand{\z}{\zeta}

\newcommand{\G}{\Gamma}

\DeclareMathOperator{\dom}{dom}

\DeclareMathOperator*{\sgn}{sgn}

\renewcommand{\Re}{\text{\rm Re}}
\renewcommand{\Im}{\text{\rm Im}}
\renewcommand{\ln}{\text{\rm ln}}

\newcommand{\no}{\notag}
\newcommand{\lb}{\label}
\newcommand{\f}{\frac}

\newcommand{\wti}{\widetilde}

\newcommand{\dott}{\,\cdot\,}

\newcommand{\bi}{\bibitem}

\let\geq\geqslant
\let\leq\leqslant

\newcommand{\al}{\a}
\newcommand{\be}{\b}

\newcommand{\ACl}{{AC_{loc}((0,\pi/2))}}

\makeatletter
\def\theequation{\@arabic\c@equation}

\allowdisplaybreaks 
\numberwithin{equation}{section}

\newtheorem{theorem}{Theorem}[section]

\theoremstyle{remark}
\newtheorem{remark}[theorem]{Remark}

\begin{document}

\title[Spectral $\zeta$-function for the Schr\"odinger operator with P\"oschl--Teller potentials]{The exotic structure of the spectral $\zeta$-function for the Schr\"odinger operator with P\"oschl--Teller potential} 

\author[G.\ Fucci]{Guglielmo Fucci}
\address{Department of Mathematics, 
East Carolina University, 331 Austin Building, East Fifth St.,
Greenville, NC 27858-4353, USA}
\email{\mailto{fuccig@ecu.edu}}
\urladdr{\url{http://myweb.ecu.edu/fuccig/}}

\author[J.\ Stanfill]{Jonathan Stanfill}
\address{Department of Mathematics, The Ohio State University \\
100 Math Tower, 231 West 18th Avenue, Columbus, OH 43210, USA}
\email{\mailto{stanfill.13@osu.edu}}
\urladdr{\url{https://u.osu.edu/stanfill-13/}}


\date{\today}

\begin{abstract}
This work focuses on the analysis of the spectral $\zeta$-function associated with a Schr\"{o}dinger operator endowed with a P\"oschl--Teller potential. We construct the spectral $\zeta$-function using a contour integral representation and, for particular self-adjoint extensions, we perform its analytic continuation to a larger region of the complex plane. We show that the spectral $\zeta$-function in these cases can possess a very unusual and remarkable structure consisting of a series of logarithmic branch points located at every nonpositive integer value of $s$ along with infinitely many additional branch points (and finitely many simple poles) whose locations depend on the parameters of the problem. By comparing the P\"oschl--Teller potential to the classic Bessel potential, we further illustrate that perturbing a given potential by a smooth potential on a finite interval can greatly affect the meromorphic structure and branch points of the spectral $\zeta$-function in surprising ways.

\end{abstract}

\maketitle




\section{Introduction and main results} \label{s1} 

The study of $\z$-functions is a topic of widespread interest in theoretical and mathematical physics, especially in the ambit of quantum field theory \cite{DW03,EORBZ94,Eli95,Ki02}. Following \cite{Sp06}, let $S$ be a sequence of complex numbers such that: i) $0\notin S$, ii) $S=\{a_{n}\}_{n\in\N}$ with $0<|a_1|\leq|a_2|\leq \cdots \to\infty$, and iii) the exponent of convergence $a_c=\lim_{n\to\infty}\textrm{sup}(\ln n)/(\ln|a_n|)$ is finite. For sequences satisfying the above conditions one can define the associated $\z$-function as the sum 
\begin{align}\label{introz}
    \zeta_{S}(s)=\sum_{n=1}^{\infty}a_{n}^{-s},
\end{align}
which converges for $\Re(s)>a_c$. Of particular importance in the field of mathematical physics is the {\it spectral} $\z$-function which is the series \eqref{introz} constructed from the spectrum $\{\lambda_n\}_{n\in\N}$ of a self-adjoint extension of a differential operator that models the dynamics of a physical system. Although the spectral $\z$-function is 
convergent for $\Re(s)$ greater than the exponent of convergence $a_\l$ of the spectral sequence $\{\lambda_n\}_{n\in\N}$ (which is finite for a wide class of problems of interest), important information about the physical system is encoded to the left of this region of the complex plane \cite{Ki02}. In order to access this information, one is, therefore, confronted with the task of analytically continuing the spectral $\z$-function to a region of the complex plane to the left of $\Re(s)=a_\l$. The process of analytic continuation of the spectral $\z$-function has been described in detail several times in the literature, most recently in \cite{FGKS21}. It consists in expressing
the spectral $\z$-function in terms of a contour integral and then using the asymptotic expansion of the integrand to obtain an analytic continuation to the left of its natural abscissa of convergence $\Re(s)=a_\l$.

The structure of the spectral $\z$-function reveals itself during the process of analytic continuation to the left of its abscissa of convergence $\Re(s)=a_\l$. This process has been performed and described in the literature for many different types of systems  and leads to a structure of the spectral $\z$-function which can be separated in two different broad categories: usual and unusual structure. 

We deem as {\it usual} the structure of a spectral $\z$-function that, in the process of analytic continuation, develops only simple poles at their ``regular" positions\footnote{In the $D$-dimensional setting, the simple poles of the spectral $\z$-function are located at $s=(D-k)/2$, for $k=\{0,\ldots,D-1\}$ and $s=-(2l+1)/2$, with $l\in\N_0$, by \cite{Ki02}.} and is never singular at the origin, such as the case of Schr\"odinger operators with smooth potentials on a finite interval (see \cite{FGKS21}) or differential operators on a smooth manifold, with or without boundary, with well-behaved coefficient functions \cite{Ki02}. This is the behavior of the spectral $\z$-function that is most widely obtained when describing a system in quantum field theory.
Any spectral $\z$-function whose structure, after analytic continuation, presents singularities other than simple poles located at their ``regular" positions is regarded as {\it unusual}. In what follows, we will list a few examples of unusual behavior that can be found in the literature. 

For instance, the spectral $\zeta$-function associated with the Laplace operator on conic manifolds exhibits an unusual structure including poles of arbitrarily high order or countably many logarithmic branch points (see, e.g., \cite{Ki08} for an outline noting, however, that  dimensions higher than one are needed to exhibit this general behavior). Other examples of non-standard spectral $\zeta$-functions, possessing poles of order higher than one and branch points, can be found,
for instance, in \cite{BKD96,CZ97,Cog06} for the generalized cone and in \cite{Ki06,FLP,FPS24} for some singular Sturm--Liouville operators. Particularly relevant for our current study are the results found in the latter three. 

In \cite{Ki06}, Kirsten, Loya, and Park studied the classic Bessel equation with potential $-(1/4)r^{-2}$ on $(0,L)$ (i.e., zero coupling constant). The authors showed that the spectral $\zeta$-function associated with the Friedrichs extension possesses a typical simple pole structure, whereas all other separated self-adjoint extensions lead to a logarithmic branch point at the origin (in addition to the simple poles from before). An important consequence of the presence of a branch point at the origin is that the functional determinant cannot be defined in the usual sense and, therefore, its definition needs to be generalized.   
The same authors then proceeded to apply their results to completely study the properties of spectral $\zeta$-functions for conical singularities in the series of papers \cite{Ki08a,Ki08b,Ki08}.

Following these efforts, Falomir, Liniado, and Pisani \cite{FLP} studied the Schr\"odinger operator with potential given by $c/r$ on $(0,L)$ for an appropriate constant $c$, comparing their results to those of the Bessel potential found in \cite{Ki06}. The authors showed that, in addition to the typical simple poles, the Friedrichs extension develops a double pole at $s=-1/2$. They further showed that other separated boundary conditions generate poles that increase in multiplicity as their distance from the origin, along the negative real axis, increases. Furthermore, it was also shown in \cite{FLP} that, while in one dimension the origin is a regular point, one can use additional dimensions to produce an operator whose spectral $\zeta$-function has a simple pole at $s=0$.

As a final example of unusual structure, we mention our recent analysis of the spectral $\zeta$-function associated with the Legendre differential equation in \cite[Sec. 5.2]{FPS24}, where we showed that the $\zeta$-function develops infinitely many simple poles located at $s=1/2$ and at the negative half-integers along with infinitely many logarithmic branch points positioned at the nonpositive integers. Due to the presence of the logarithmic branch point at the origin, the functional determinant is not defined, and one must introduce a regularized version of it, just like the one provided in \cite[Sec. 1]{Ki06}. See also Subsection \ref{unusualstructure} where we relate the Legendre equation to this study.

This study should be considered an application of our recent work \cite{FPS24}. In particular, applying those results to a Schr\"{o}dinger equation endowed with a P\"oschl--Teller potential, we illustrate that the associated spectral $\z$-function can possess an unusual and remarkable structure, consisting of simple poles whose positions and number depend on the parameter choices and self-adjoint extension, and of logarithmic branch points located at every nonpositive integer value of $s$ along with, possibly, infinitely many additional logarithmic branch points whose locations depend, once again, on the parameters of the problem.
We also compare these results to the classic Bessel equation to show that the structure of the $\zeta$-function can be greatly affected by perturbing a given potential by a smooth potential.

We would like to underscore the importance of the P\"oschl--Teller system considered in this work with regard to the structure of its spectral $\z$-function. In particular, spectral $\z$-functions with countably many logarithmic branch points are rarely encountered in the literature, let alone one with additional poles and infinitely many more logarithmic branch points whose locations depend on the parameters of the operator. This is the first such example we have encountered, and this paper aims to add one more item to the very short list of systems possessing a spectral $\z$-function with an unusual structure.
 
We now turn our attention to the description of the operators that are the main focus of our project.

\subsection{P\"oschl--Teller potential and spectral \texorpdfstring{$\zeta$}{zeta}-function}

We study Sturm--Liouville operators associated with the differential expression with P\"oschl--Teller potential \cite{PT33} given by
\begin{align}
\tau_{\mu,\nu} = - \dfrac{d^2}{dx^2} +\dfrac{\mu^2-(1/4)}{\sin^2 (x)}+\dfrac{\nu^2-(1/4)}{\cos^2 (x)}, \quad \mu,\nu \in [0,\infty), \; x\in(0,\pi/2),    \label{2.1} 
\end{align}
which is in the limit circle case at the endpoint $x=0$ if and only if $\mu\in[0,1)$, and in the limit circle case at $x=\pi/2$ if and only if 
$\nu \in [0,1)$. In addition, \eqref{2.1} is nonoscillatory for all $\mu,\nu\in[0,\infty)$ at both endpoints.
Thus, one can conclude that for $\mu,\nu \in [0,1)$, \eqref{2.1} is quasi-regular on $x\in(0,\pi/2)$ and the associated minimal operator is bounded from below.

The exact solvability of one-dimensional Schr\"odinger equations with potentials given as in \eqref{2.1} (and its hyperbolic analog) were first illustrated by Rosen and Morse \cite{RM32} in 1932, P\"oschl and Teller \cite{PT33} in 1933, and Lotmar \cite{Lo35} in 1935, finding many applications thereafter. Since these initial studies, problems of this type were discussed by Infeld and Hull \cite{IH51}, with the singular periodic problems presented by Scarf \cite{Sc58}, \cite{GK85}, and \cite{GMS85}. More recently, these types of problems have also been summarized in  \cite[Sec.~38, 39, 93]{Fl99} and \cite{DW11}, as well as applied in \cite{GPS21} to prove a refinement of the classic Hardy's inequality.

An overview of the basic notions of singular Weyl--Titchmarsh--Kodaira theory that are necessary for a description of all self-adjoint extensions of
the minimal quasi-regular Sturm--Liouville operator can be found, for instance, in \cite[Ch.~4, 6--8]{Ze05}, \cite{GLN20}, \cite[Ch.~13]{GNZ23}, and also the summary in \cite[App. A]{FPS24}. Here, for the sake of conciseness, we limit ourselves to only stating the
indispensable definitions and theorems, and invite the reader interested in the details to consult the references mentioned above.  
Given $\tau_{\mu,\nu}$ as in \eqref{2.1}, the \textit{maximal operator} $T_{\mu,\nu,max}$ in $L^2((0,\pi/2);dx)$ associated with $\tau_{\mu,\nu}$ is defined by
\begin{align}
&T_{\mu,\nu,max} f = \tau_{\mu,\nu} f,    
\\
& f \in \dom(T_{\mu,\nu,max})=\big\{g\in L^2((0,\pi/2);dx) \, \big| \,g,g'\in\ACl;    
\tau g\in L^2((0,\pi/2);dx)\big\},  \notag
\end{align}
while the \textit{minimal operator} $T_{\mu,\nu,min}$ in $L^2((0,\pi/2);dx)$ associated with $\tau_{\mu,\nu}$ is
\begin{align}
&T_{\mu,\nu,min} f = \tau_{\mu,\nu} f, 
\\
&f \in \dom(T_{\mu,\nu,min})=\big\{g\in\dom(T_{\mu,\nu,max})  \, \big| \, W(h,g)(0)=0=W(h,g)(\pi/2) \, 
\text{for all } h\in\dom(T_{\mu,\nu,max}) \big\},   \no 
\end{align}
where, for $f,g\in\ACl$, 
\begin{equation}
W(f,g)(x) = f(x)g'(x) - f'(x)g(x), \quad x \in (0,\pi/2), 
\end{equation}
is the Wronskian of $f$ and $g$.
In order to characterize the self-adjoint extensions associated with \eqref{2.1} we need to introduce the generalized boundary value of a function $g\in\dom(T_{\mu,\nu,max})$. 
 
It is important to note that the generalized boundary values depend only on the leading behavior of \eqref{2.1} near the singular endpoint. As $x\to 0$, the differential expression $\tau_{\mu,\nu}$ reduces to a bounded perturbation of the Liouville form of the Bessel differential expression (see, e.g., \cite[Sec. 12]{Ev05}). A similar observation holds as one approaches the right endpoint, $\pi/2$. These considerations show that one can follow the analysis outlined in \cite[Eqs. (2.27)--(2.30)]{GPS24} (see also \cite[App. A]{GPS21}) to conclude that the generalized boundary values for $g\in\dom(T_{\mu,\nu,max})$ are of the form
\begin{align}
\wti g(0)&=\begin{cases}
\lim_{x\downarrow0}g(x)/\big[(2\mu)^{-1}x^{(1-2\mu)/2}\big], & \mu\in(0,1),\\[1mm]
\lim_{x\downarrow0}g(x)/\big[-x^{1/2}\ln(x)\big], & \mu=0,
\end{cases}    \label{2.2} \\[1mm]
\wti g^{\, \prime}(0)&=\begin{cases}
\lim_{x\downarrow0}\big[g(x)-\wti g(0)(2\mu)^{-1}x^{(1-2\mu)/2}\big]\big/x^{(1+2\mu)/2}, & \mu\in(0,1),\\[1mm]
\lim_{x\downarrow0}\big[g(x)+\wti g(0)x^{1/2}\ln(x)\big]\big/x^{1/2}, & \mu=0,
\end{cases}    \label{2.3} \\[1mm]
\wti g(\tfrac{\pi}{2})&=\begin{cases}
\lim_{x\uparrow\tfrac{\pi}{2}}g(x)/\big[(2\nu)^{-1}(\tfrac{\pi}{2}-x)^{(1-2\nu)/2}\big], & \nu\in(0,1),\\[1mm]
\lim_{x\uparrow\tfrac{\pi}{2}}g(x)/\big[-(\tfrac{\pi}{2}-x)^{1/2}\ln(\tfrac{\pi}{2}-x)\big], & \nu=0,
\end{cases}     \label{2.4} \\[1mm]
\wti g^{\, \prime}(\tfrac{\pi}{2})&=\begin{cases}
\lim_{x\uparrow\tfrac{\pi}{2}}\big[g(x)-\wti g(\tfrac{\pi}{2})(2\nu)^{-1}(\tfrac{\pi}{2}-x)^{(1-2\nu)/2}\big]\big/\big[-(\tfrac{\pi}{2}-x)^{(1+2\nu)/2}\big], & \nu\in(0,1),\\[1mm]
\lim_{x\uparrow\tfrac{\pi}{2}}\big[g(x)+\wti g(\tfrac{\pi}{2})(\tfrac{\pi}{2}-x)^{1/2}\ln(\tfrac{\pi}{2}-x)\big]\big/\big[-(\tfrac{\pi}{2}-x)^{1/2}\big], & \nu=0.
\end{cases}    \label{2.5} 
\end{align}
We would like to point out that the constants in \eqref{2.2}--\eqref{2.5} have been chosen such that when $\mu=1/2$ (resp., $\nu=1/2$) the generalized boundary values agree with the regular boundary values at $x=0$ (resp., $x=\pi/2$) for $g\in\dom(T_{\mu,\nu,max})$.

The generalized boundary values \eqref{2.2}--\eqref{2.5} are used to represent all the self-adjoint extensions of $T_{\mu,\nu,min}$. More precisely, according to the theory of singular Sturm--Liouville operators, all the self-adjoint extensions of $T_{\mu,\nu,min}$ can be divided in two classes: the first class is defined by imposing, on $g\in\dom(T_{\mu,\nu,max})$, \emph{separated boundary conditions} 
\begin{align}
& T_{\mu,\nu}^{(\al,\be)} f = \tau f, \quad \al,\be\in[0,\pi),   \notag\\
& f \in \dom(T_{\mu,\nu}^{(\al,\be)})=\big\{g\in\dom(T_{\mu,\nu,max}) \, \big| \, \wti g(0)\cos(\al)+ {\wti g}^{\, \prime}(0)\sin(\al)=0;   \label{2.12text} \\ 
& \hspace*{6.15cm} \, \wti g(\pi/2)\cos(\be)- {\wti g}^{\, \prime}(\pi/2)\sin(\be) = 0 \big\},    \notag
\end{align}
while the second class is defined by imposing, on $g\in\dom(T_{\mu,\nu,max})$, \emph{coupled boundary conditions} 
\begin{align}
\begin{split}\label{teq0} 
& T_{\mu,\nu}^{(\varphi,R)} f = \tau f,    \\
& f \in \dom(T_{\mu,\nu}^{(\varphi,R)})=\bigg\{g\in\dom(T_{\mu,\nu,max}) \, \bigg| \begin{pmatrix} \wti g(\pi/2)\\ {\wti g}^{\, \prime}(\pi/2)\end{pmatrix} 
= e^{i\varphi}R \begin{pmatrix}
\wti g(0)\\ {\wti g}^{\, \prime}(0)\end{pmatrix} \bigg\},
\end{split}
\end{align}
where $\varphi\in[0,\pi)$, and $R \in SL(2,\bbR)$. We will, henceforth, collectively denote by $T_{\mu,\nu}^{(A,B)}$ both the separated and coupled self-adjoint extensions. 

The construction of the spectral $\zeta$-function associated with the self-adjoint extensions $T_{\mu,\nu}^{(A,B)}$ of the quasi-regular Sturm--Liouville operator \eqref{2.1} has been provided in detail, for general quasi-regular Sturm--Liouville operators, in our work \cite{FPS24}. In particular, the spectral $\zeta$-function associated with $T_{\mu,\nu}^{(A,B)}$ is defined as
\begin{align}
\zeta(s;T_{\mu,\nu}^{(A,B)}):=\sum_{\underset{\lambda_j\neq 0}{j\in J}} \lambda^{-s}_{\mu,\nu,j}(A,B),
\end{align}
with $\Re(s)>1/2$, and can be represented in terms of a contour integral as (\cite[Sec. 3]{FPS24})
\begin{equation}\label{16}
\zeta(s;T_{\mu,\nu}^{(A,B)})=\frac{1}{2\pi i}\int_{\gamma}dz\,z^{-s}\left[\frac{d}{dz}\ln\,F_{\mu,\nu}^{(A,B)}(z)-z^{-1}m_{0}\right].
\end{equation} 
In this expression,  $\gamma$ represents a contour in the complex plane that encircles the spectrum $\sigma(T_{\mu,\nu}^{(A,B)})$ in a counterclockwise direction while avoiding the origin and $m_{0}=m(0,T_{\mu,\nu}^{(A,B)})$ is the multiplicity of the zero eigenvalue. Because of the presence of $z^{-s}$ we choose, following \cite{KM03} (see also \cite{KM04}), the branch cut for the definition of the integral to be 
\begin{align}\label{1a}
    R_\psi=\{z=te^{i\psi}:t\in [0,\infty)\},\quad \psi\in (\pi/2,\pi).
\end{align}
$F_{\mu,\nu}^{(A,B)}(z)$, in the integrand of \eqref{16}, is the \textit{characteristic function} of the self-adjoint extension $T_{\mu,\nu}^{(A,B)}$ and is an entire function of $z\in\C$ of order $1/2$ with isolated zeros coinciding with the eigenvalues of $T_{\mu,\nu}^{(A,B)}$ \cite[Sec. 3]{FPS24}. The construction of the characteristic function for the operators studied here is discussed in Section \ref{s2} and the analytic continuation of \eqref{16} for certain self-adjoint extensions is described in Section \ref{s3}. We now turn to our main results.

\subsection{The unusual structure of the spectral \texorpdfstring{$\zeta$}{zeta}-function associated with certain extensions}\lb{unusualstructure}

The main purpose of the current investigation is to highlight the unusual structure that the spectral $\zeta$-function acquires in the process of analytic continuation. Since the structure of the $\zeta$-function heavily depends on the self-adjoint extension chosen, we focus on those particular extensions that produce interesting and atypical behaviors of the associated $\zeta$-function. 
Due to the complicated nature of the asymptotic expansions encountered in the analysis, some cases require restrictions on parameter choices.
The remarkable structure of the spectral $\zeta$-function associated with these extensions is summarized in the following theorem, with the proof given in Section \ref{s3} (see also \eqref{2.24}--\eqref{2.25b}).
\begin{theorem}\label{thmstructure}
Let $T_{\mu,\nu}^{(\alpha,\beta)}$ denote the self-adjoint extension of the separated boundary condition for the minimal operator associated with \eqref{2.1}. Then the following $(i)$--$(iv)$ hold$:$

\begin{itemize}
\item[$(i)$] The spectral $\zeta$-functions associated with the Friedrichs, Friedrichs--Neumann-type, Neumann--Friedrichs-type, and Neumann-type extensions are respectively given by
\begin{align}\label{2.37}
\zeta(s;T_{\mu,\nu}^{(0,0)})&=2^{-2s}\zeta_{H}\left(2s,\tfrac{1+\mu+\nu}{2}\right),\quad \mu,\nu\in[0,1),\\
\zeta(s;T_{\mu,\nu}^{(0,\pi/2)})&=2^{-2s}\zeta_{H}\left(2s,\tfrac{1+\mu-\nu}{2}\right),\quad \mu\in[0,1),\; \nu\in(0,1),\\
\zeta(s;T_{\mu,\nu}^{(\pi/2,0)})&=2^{-2s}\zeta_{H}\left(2s,\tfrac{1-\mu+\nu}{2}\right),\quad \mu\in(0,1),\; \nu\in[0,1),\\
\zeta(s;T_{\mu,\nu}^{(\pi/2,\pi/2)})&=\begin{cases}2^{-2s}\zeta_{H}\left(2s,\frac{1-\mu-\nu}{2}\right),& \mu+\nu\neq 1,\\
2^{-2s}\zeta_{R}(2s),& \mu+\nu= 1,
\end{cases}\quad \mu,\nu\in(0,1),
\end{align}
where $\zeta_R$ and $\zeta_H$ denote the Riemann and Hurwitz $\zeta$-functions, respectively. In particular, each spectral $\zeta$-function can always be extended to a meromorphic function of $s$ with only one simple pole at $s=1/2$.
\item[$(ii)$] The spectral $\zeta$-function associated with the extensions $T_{0,\nu}^{(\a,\b)}$, $\nu\in(0,1)$, $\a\in(0,\pi)$, $\b\in\{0,\pi/2\}$, has a simple pole at $s=1/2$ and logarithmic branch points at $s=-j,\ j\in\bbN_0=\bbN\cup\{0\}$.
\end{itemize}
\noindent Let $\nu=p/q$ with $p,q\in\N$, $0< p\leq q-1$, and $p,q$ relatively prime. Then, generically,\footnote{See Sections \ref{sub3.1.1}, \ref{sub3.1.2}, and Remark \ref{RemB3}$)$ for the exact notion of \textit{generic} in this setting which is in reference to whether certain terms of the asymptotic expansion corresponding to the poles and branch points are nonzero.} one has the following$:$

\begin{itemize}
    \item[$(iii)$] The spectral $\zeta$-function associated with the extensions $T_{0,p/q}^{(0,\b)}$, $\b\in(0,\pi)\backslash\{\pi/2\}$, has simple poles at $s=1/2$, at $s=-mp/q$ with $1\leq m< q$, and at $s=-(M+p)/q$ with $M\geq q$, $M\neq jp$ for $0\leq j\leq q-1$, and $M\neq kq-p$ for $k\in \bbN$.
    \item[$(iv)$] The spectral $\zeta$-function associated with the extensions $T_{0,p/q}^{(\a,\b)}$, $\a\in(0,\pi)$, $\b\in(0,\pi)\backslash\{\pi/2\}$, has simple poles at $s=1/2$ and at $s=-mp/q$ with $1\leq m< q$, as well as logarithmic branch points at $s=-j,\ j\in\bbN_0=\bbN\cup\{0\}$ and at $s=-(M+p)/q$ with $M\geq q$ and $M\neq jp$ for $0\leq j\leq q-1$.
\end{itemize}
\end{theorem}

In the rest of this subsection we would like to offer some remarks regarding the ways in which a perturbation of the potential of a Sturm--Liouville operator can change the structure of the associated spectral $\z$-function.

We can start by focusing on the simplest of perturbations, namely a shift in the potential by a small constant. We would like to point out that a constant positive shift in the potential can be interpreted, in the ambit of physical systems, as the addition of a mass to an otherwise massless field. In what follows, we also include a scaling factor which will be useful in the explicit example we illustrate later. First, denote by $\{\lambda_n\}_{n\in J},\ \lambda_{j-1}<\lambda_j,$ the spectrum of a nonnegative Sturm--Liouville operator with $\lambda_0$ being the zero eigenvalue of multiplicity $m_0\in\{1,2\}$, if it exists, so that $J=\bbN_0$ if $\lambda_0$ exists and $J=\bbN$ otherwise. Next, denote by $\{\mu_n\}_{n\in J}$ the spectrum of another operator that satisfies the relation $\mu_n=\alpha_1\lambda_n+\alpha_2$ with $\a_1>0,\ \a_2\in\bbR\backslash\{0\},$ such that  $|\a_2|<\a_1\l_1$. When $\a_1=1$, the second operator can be understood as simply the first with the potential shifted by $\a_2$. By using a Mellin transform one can write 
\begin{align}
    \mu_{n}^{-s}=\a_1^{-s}\sum_{k=0}^{\infty}\frac{(-1)^{k}}{k!}\left(\frac{\a_2}{\a_1}\right)^k\frac{\Gamma(s+k)}{\Gamma(s)}\lambda_{n}^{-s-k},\quad n\in\N,\ \Re(s)>0.
\end{align}
Fubini's theorem can then be used to obtain the following relation between the spectral $\z$-function, $\z_{\a_1,\a_2}(s)$, associated with the spectrum $\{\mu_n\}_{n\in J}$ and $\z_{0}(s)$ associated with the spectrum $\{\lambda_n\}_{n\in J}$ valid for $\Re(s)>1/2$ and $m_0\in\{0,1,2\}$ (where $m_0=0$ denotes no zero eigenvalue $\l_0$),
\begin{align}\lb{zetarelation}
  \z_{\a_1,\a_2}(s)=m_0\a_2^{-s}+\a_1^{-s}\sum_{k=0}^{\infty}\frac{(-1)^{k}}{k!}\left(\frac{\a_2}{\a_1}\right)^k\frac{\Gamma(s+k)}{\Gamma(s)}\z_{0}(s+k),  
\end{align}
and by analytic continuation elsewhere.

The formula \eqref{zetarelation} can be used to compare the singularity structure of $\z_{\a_1,\a_2}(s)$, associated with the operator with shifted potential, to the one of $\z_0(s)$ corresponding to the original, non-shifted, operator. For instance, it is clear from \eqref{zetarelation} that if $\z_0(s)$ has a simple pole only at $s=1/2$, then the spectral $\z$-function of the shifted operator develops infinitely many simple poles at the usual points $s=-(2n-1)/2$, with $n\in\N_{0}$. More surprisingly, if $\z_0(s)$ has simple poles at the usual points, namely $s=-(2n-1)/2$, with $n\in\N_{0}$, then $\z_{\a_1,\a_2}(s)$ {\it may of may not} inherit simple poles at those usual locations. In fact, an induction argument and the relation \eqref{zetarelation} can be used to prove that if the following holds 
\begin{align}\lb{residuerelation}
    \textrm{Res}\,\z_{0}\left(\frac{1}{2}-k\right)=\frac{\sqrt{\pi}}{k!\Gamma(1/2-k)} \left(\frac{\a_2}{\a_1}\right)^k\textrm{Res}\,\z_{0}\left(\frac{1}{2}\right),\quad k\in\N_0,
\end{align}
then the spectral $\z$-function $\z_{\a_1,\a_2}(s)$ will develop a simple pole at $s=1/2$ and nowhere else.

An example of the spectral shift discussed above in the ambit of the operators of interest in this work can be found by comparing the Legendre expression $\tau_{Leg} = - (d/dX) (1 - X^2) (d/dX)$, $X \in (-1,1)$, recently studied in \cite[Sec. 5.2]{FPS24}, to the P\"oschl--Teller one.  
We begin with the more general Jacobi differential expression (see the recent thorough spectral study \cite{GLPS23}) 
\begin{align}\lb{Jacobi}
\begin{split} 
\tau^{Jac}_{\d,\eta} = - (1-X)^{-\d} (1+X)^{-\eta}(d/dX) \big((1-X)^{\d+1}(1+X)^{\eta+1}\big) (d/dX),\quad \d, \eta \in \bbR, \; X \in (-1,1).
\end{split} 
\end{align}
Notice that the Legendre expression is realized by choosing $\d=\eta=0$ in \eqref{Jacobi}. Furthermore, by \cite[Sec. 24]{Ev05}\footnote{We point out that the Liouville normal form given at the end of \cite[Sec. 19]{Ev05} has the incorrect sign on the potential.}, the Liouville normal form (or Schr\"odinger form) of the Jacobi equation $\tau^{Jac}_{\d,\eta}y=Zy$ is given by (note the shift in the spectral parameter)
\begin{align}\label{normJac}
-\dfrac{d^2}{d\xi^2}v(Z,\xi)+\frac{\d^2-(1/4)}{4\sin^2((2\xi-\pi)/4)}v(Z,\xi)+\frac{\eta^2-(1/4)}{4\sin^2((2\xi+\pi)/4)}&v(Z,\xi)=\big[Z+((\d+\eta+1)/2)^2\big]v(Z,\xi),\no\\
\d,\;& \eta\in\bbR,\; \xi\in(-\pi/2,\pi/2),\; Z\in\bbC.
\end{align}
Utilizing the simple change of variables $\xi=2x-(\pi/2)$ recasts \eqref{normJac} into the problem
\begin{align}\label{transformnormJac}
\begin{split}
-\frac{d^2}{dx^2}u(Z,x)+\frac{\eta^2-(1/4)}{\sin^2(x)}u(Z,x)+\frac{\delta^2-(1/4)}{\cos^2(x)}u(Z,x)=\big[4Z+(\d+\eta+1)^2\big] u(Z,x),\\
\d, \eta\in\bbR,\; x\in(0,\pi/2),\; Z\in\bbC.
\end{split}
\end{align}
Let us point out that while \eqref{2.1} remains invariant under a change of sign in $\mu$ and $\nu$, a change of sign in $\d$ and $\eta$ in \eqref{transformnormJac} corresponds, instead, to subtracting a different constant from the potential or, equivalently, to performing a different spectral shift.

Thus \eqref{transformnormJac} shows that studying the original Jacobi equation $\tau^{Jac}_{\d,\eta}y=Zy$ is equivalent to studying the problem $\tau_{\mu,\nu}u=zu$ with the spectral parameters related via the equation $z=4Z+(\sgn(\d)\nu+\sgn(\eta)\mu+1)^2$ where $\sgn(\dott)$ is the sign function (i.e., $\d=\sgn(\d)\nu$ and $\eta=\sgn(\eta)\mu$). Notice that this change in the spectral parameter is exactly the square of the expression $\sigma_{\d,\eta}(z)$ found in \cite[Eq. (5.5)]{GLPS23}. 

In particular, the Legendre equation $\tau_{Leg}y=Zy$ studied in \cite[Sec. 5.2]{FPS24} is equivalent to the P\"oschl--Teller equation with $\mu=\nu=0$, that is, $\tau_{0,0}u=zu$, with the spectral parameters related through the equation $z=4Z+1$.  This implies that the spectral $\z$-function associated with a given self-adjoint extension of $T_{0,0,min}$ can be obtained through \eqref{zetarelation} from the spectral $\z$-function associated to the same self-adjoint extension of $T_{Leg,min}$. If, for instance, we consider the Friedrichs extension then the eigenvalues of $T_{0,0}^{(0,0)}$ are (c.f. \eqref{2.37}) $\mu_n=(2n+1)^{2}$, with $n\in\N_0$ and the associated spectral $\z$-function is given in \eqref{2.37} in terms of the Hurwitz $\zeta$-function. Since $z=4Z+1$, then the eigenvalues of the Friedrichs extension of the Legendre operator, $T_{Leg}^{(0,0)}$, can be found to be $\lambda_k=k(k+1)$, with $k\in\N_0$ and its spectral $\zeta$-function is a one-dimensional Epstein $\zeta$-function
\begin{equation}\lb{legendrezeta}
\z(s;T_{Leg}^{(0,0)})=\sum_{k\in\bbN}\big(k^2+k\big)^{-s},\quad \Re(s)>1/2.
\end{equation}
By \eqref{zetarelation} with $\a_1=4,\ \a_2=1,\ m_0=1$, one obtains
\begin{align}\lb{legendrezeta1}
    \z(s;T_{0,0}^{(0,0)})=2^{-2s}\zeta_{H}(2s,1/2)=1+4^{-s}\sum_{k=0}^{\infty} \frac{(-1)^{k}}{4^{k}k!}\frac{\Gamma(s+k)}{\Gamma(s)}\z(s+k;T_{Leg}^{(0,0)}).
\end{align}
The pole structure of the one-dimensional Epstein $\zeta$-function in \eqref{legendrezeta} can be found by utilizing \cite[Eq. (6.7)]{Eli95} with $a=1$ and $b=c=0$, in particular one can show that for $j\in\N_0$
\begin{align}\lb{residuelegendre}
   \textrm{Res}\,\z\left(\frac{1}{2}-j;T_{Leg}^{(0,0)}\right)= \frac{\sqrt{\pi}}{4^{j}j!\Gamma(1/2-j)}\left(\frac{1}{2}\right).
\end{align}
The residue of $\z(s;T_{Leg}^{(0,0)})$ at the negative half-integers and at $s=1/2$ given in \eqref{residuelegendre} matches the form given in \eqref{residuerelation} with $\a_2/\a_1=1/4$ (since $\textrm{Res}\,\z(1/2;T_{Leg}^{(0,0)})=1/2$) which implies that the spectral $\z$-function of $T_{0,0}^{(0,0)}$ on the left-hand side of \eqref{legendrezeta1} develops only a simple pole at $s=1/2$, as expected. This represents an explicit example in which the constant shift of the spectrum causes the resulting spectral $\zeta$-function to ``lose" its poles at the negative half-integers.

Similar modifications to the structure of the spectral $\z$-function can occur when a given potential is modified by a smooth, bounded perturbation as we illustrate below in the ambit of the Schr\"odinger equation with a P\"oschl--Teller potential.   
First notice that
\begin{equation}
-\frac{1}{4\sin^2(x)}=\sum_{n\in\bbN_0}\frac{(-1)^{n}4^{n-1}(2n-1)B_{2n} x^{2n-2}}{(2n)!},\quad 0<|x|<\pi,
\end{equation}
which follows from taking the derivative of the series expansion for $\cot(x)$ (see, e.g., \cite[Eq. 4.19.6]{DLMF}).
This shows that the differential operators studied here can be understood as a perturbation of the typical Bessel operator with coupling constant zero. In particular, the case $\mu=0,\ \nu=1/2,$ in \eqref{2.1} shows that perturbing the Bessel expression with potential $q_0(x)=-(1/4)x^{-2}$ on $(0,\pi/2)$ by
\begin{equation}\label{perturb}
q_{per}(x)=\sum_{n\in\bbN}\frac{(-1)^{n}4^{n-1}(2n-1)B_{2n} x^{2n-2}}{(2n)!},
\end{equation}
which is smooth on $(0,\pi)$, can affect the structure of the spectral $\zeta$-function for different self-adjoint extensions in wildly different ways! 

For instance, the spectral $\zeta$-function associated with the Friedrichs extension of the Bessel operator with potential $q_0(x)$ on $(0,\pi/2)$ has simple poles in $s$ at all negative half-integers in addition to the typical simple pole at $s=1/2$ (cf. \cite[Thm. 1.1 (3)]{Ki06}). 
However, perturbing the potential by $q_{per}(x)$ yields the differential expression defined by \eqref{2.1} with $\mu=0$ and $\nu=1/2$, for which \eqref{2.37} shows that the corresponding spectral $\zeta$-function has only one simple pole at $s=1/2$. Hence, the addition of $q_{per}(x)$ to the Bessel operator with potential $q_0(x)$ ``simplifies" the meromorphic structure of the associated spectral $\zeta$-function by removing infinitely many simple poles (a phenomenon that, as we have shown earlier in this discussion, occurs also for simple constant shifts).

On the other hand, when considering any self-adjoint extension defined by $\b=0$ and a boundary condition other than Friedrichs at $x=0$ (i.e., $\alpha\neq0$ in \eqref{2.12text}) of the Bessel operator with potential $q_0(x)$ on $(0,\pi/2)$, the spectral $\zeta$-function has, in addition to the simple poles at $s=-(2n-1)/2,\ n\in\bbN_0$, a logarithmic branch point at $s=0$ as proven in \cite[Thm. 1.2 (3)]{Ki06} (see also the generalization in \cite[Sec. 5.1]{FPS24}). Remarkably, by Theorem \ref{thmstructure} $(ii)$, the additional terms of $q_{per}(x)$ modify the structure in two ways: First, they cause the spectral $\zeta$-function to develop logarithmic branch points at all negative integers and, second, they remove the simple poles at the negative half-integers of the corresponding Bessel operator's $\zeta$-function. In this case, the additional terms of the potential $q_{per}(x)$ ``worsen" the structure of the spectral $\z$-function by adding infinitely many branch points. 
We would like to point out that the same situation can occur when the potential is modified by a simple constant shift. In fact, the relation \eqref{zetarelation} shows that if the spectral $\z$-function $\z_0(s)$ of the unperturbed operator possesses a logarithmic branch point at the origin, then the spectral $\z$-function $\z_{\a_1,\a_2}(s)$ of the operator with the potential shifted by a constant develops logarithmic branch points at all negative integers.  

This shows that modifying a given potential by a smooth perturbation on a finite interval can drastically change the structure of the associated spectral $\zeta$-functions for better or worse depending on the given self-adjoint extension, even though a smooth potential is typically associated with only possible simple poles at $s=-(2n-1)/2,\ n\in\bbN_0$ (see \cite[Sec. 3]{FGKS21}).

\subsection{The \texorpdfstring{$\zeta$}{zeta}-regularized functional determinant}

According to the general theory \cite{RS71}, one defines the $\z$-regularized functional determinant of a suitable operator $T$ in terms of its spectral $\z$-function $\zeta(s;T)$ as
\begin{align}\lb{zetadet}
    \det(T)=\exp(-\tfrac{d}{ds}\zeta(s;T)|_{s=0}).
\end{align}
According to Theorem \ref{thmstructure}, the spectral $\zeta$-function associated with self-adjoint extensions characterized by $\a \neq 0$ exhibits a logarithmic branch point at the origin, resulting in the functional determinant in \eqref{zetadet} being undefined. In order to overcome this problem, we follow \cite{Ki06} and utilize a notion of regularized determinant that is appropriate when a logarithmic branch point is present at the origin. More precisely, we introduce the modified $\zeta$-function
\begin{equation}\label{modzeta}
\zeta_{\textrm{reg}}(s;T_{0,\nu}^{(\alpha,\beta)}):=\zeta(s;T_{0,\nu}^{(\alpha,\beta)})+s\,\ln s,\quad \a\in(0,\pi),\; \b\in[0,\pi),\; \nu\in(0,1),
\end{equation}
which is obtained by subtracting from \eqref{3.12} and \eqref{3.17} their singular $s\to 0$ behavior (cf. \eqref{4.4}),  and which has, consequently, a well-defined derivative at $s=0$. One can then define the regularized determinant in terms of \eqref{modzeta} as
\begin{equation}\label{moddet}
{\det}_{\textrm{reg}}(T_{0,\nu}^{(\alpha,\beta)}):=\exp(-\tfrac{d}{ds}\zeta_{\textrm{reg}}(s;T)|_{s=0}).
\end{equation}
We now state the analog of \cite[Thm. 1.6]{Ki06} for the operators considered here with the proof given in Section \ref{s4} following the analysis in \cite[Sec. 9]{Ki06} (see also \eqref{2.29}).

\begin{theorem}\label{thmdet}
Define the regularized determinant by \eqref{moddet}. Then, for $\nu\in(0,1)$,
\begin{align}
{\det}_{\textup{reg}}(T_{0,\nu}^{(\alpha,\beta)})&=\begin{cases}
\exp\bigg(\ln\big(\mathcal{F}^{(\a,\b)}_{0,\nu; m_0}\big)-\ln\left[-\dfrac{\Gamma(-\nu)\sin(\a)\sin(\b)}{2^{\nu+2}\pi}\right]+i\pi m_0+\gamma_E\bigg),& \alpha,\beta\in(0,\pi),\\[4mm]
\exp\bigg(\ln\big(\mathcal{F}^{(\a,0)}_{0,\nu; m_0}\big)-\ln\left[-\dfrac{\Gamma(1+\nu)\sin(\a)2^{\nu-1}}{\pi}\right]+i\pi m_0+\gamma_E\bigg),& \a\in(0,\pi),\; \b=0,
\end{cases}\\
{\det}(T_{0,\nu}^{(0,\beta)})&=\begin{cases}
\exp\bigg(\ln\big(\mathcal{F}^{(0,\b)}_{0,\nu; m_0}\big)-\ln\left[\dfrac{\Gamma(-\nu)\sin(\b)}{2^{\nu+1}\pi}\right]+i\pi m_0\bigg),& \beta\in(0,\pi),\\[4mm]
\exp\bigg(-2\ln\left[\Gamma\left(\dfrac{1+\nu}{2}\right)\right]+\ln (2^{1-\nu}\pi)\bigg),& \b=0,
\end{cases}
\end{align}
where $m_0\in\{0,1\}$ denotes the multiplicity of zero as an eigenvalue and, for $F_{0,\nu}^{(\a,\b)}(\dott)$ given in \eqref{3.2}, we define 
\begin{align}
\lim_{z\downarrow 0}\,(z)^{-m_0}F_{0,\nu}^{(\a,\b)}\left(z\right)=\mathcal{F}^{(\a,\b)}_{0,\nu; m_0}\neq 0.  
\end{align}
\end{theorem}

\subsection{Structure of paper}
The remainder of the paper is structured as follows. In Section \ref{s2}, we construct the characteristic function $F_{A,B}(z)$ by introducing a fundamental system of solutions of $\tau_{\mu,\nu} u=zu$, $z \in \bbC$, normalized at $x=0$. We then discuss certain self-adjoint extensions that admit an explicit spectrum, allowing their spectral $\zeta$-function to be related to other, well-known, $\zeta$-functions, such as the Riemann and Hurwitz $\zeta$-functions. Next, we explicitly perform the analytic continuation of the spectral $\zeta$-function in Section \ref{s3}, while Section \ref{s4} contains the proof of Theorem \ref{thmdet} regarding the $\zeta$-regularized functional determinant. The appendices include general results that complement the rest of the paper, such as the connection between the hypergeometric differential equation and the Schr\"odinger equation with P\"oschl--Teller potential in Appendix \ref{sA} and certain needed asymptotic expansions in Appendix \ref{sB}.

\section{The characteristic function and some extensions with explicit spectrum}\label{s2}

In order to provide an explicit expression for the characteristic function we need to introduce the normalized (at $x=0$) fundamental system of solutions $\phi_{\mu,\nu}(z,\dott,0)$ and 
$\theta_{\mu,\nu}(z,\dott,0)$ of $\tau_{\mu,\nu} u=zu$, $z \in \bbC$. These solutions are defined by imposing the generalized boundary conditions 
\begin{align}\label{2.6}
\wti \theta_{\mu,\nu}(z,0,0)=1,\quad \wti \theta^{\, \prime}_{\mu,\nu}(z,0,0)=0,  \quad 
\wti \phi_{\mu,\nu}(z,0,0)=0,\quad \wti \phi^{\, \prime}_{\mu,\nu}(z,0,0)=1, 
\end{align} 
with $\phi_{\mu,\nu}(\dott,x,0),\ \theta_{\mu,\nu}(\dott,x,0)$ entire for fixed $x \in (0,\pi/2)$ and are given by (see Appendix \ref{sA})
\begin{align}
\begin{split}\label{2.7}
\phi_{\mu,\nu}(z,x,0)&=[\sin(x)]^{(1+2\mu)/2}[\cos(x)]^{(1+2\nu)/2}\\
&\quad\times\mathstrut_2F_1\big(\big(1+\mu+\nu+z^{1/2}\big)/2,\big(1+\mu+\nu-z^{1/2}\big)/2;1+\mu;\sin^2(x)\big),\quad \mu\in[0,1),
\end{split}\\
\begin{split}\notag
\theta_{\mu,\nu}(z,x,0)&=\begin{cases}
(2\mu)^{-1}[\sin(x)]^{(1-2\mu)/2}[\cos(x)]^{(1-2\nu)/2}\\
\quad \times\mathstrut_2F_1\big(\big(1-\mu-\nu+z^{1/2}\big)/2,\big(1-\mu-\nu-z^{1/2}\big)/2;1-\mu;\sin^2(x)\big), & \hspace{-.7cm} \mu\in(0,1),\\[2mm]
-2^{-1}[\sin(x)]^{1/2}[\cos(x)]^{(1+2\nu)/2}\\
\ \times \bigg\{\mathstrut_2F_1\big(\big(1+\nu+z^{1/2}\big)/2,\big(1+\nu-z^{1/2}\big)/2;1;\sin^2(x)\big) \, \ln\big(\sin^2(x)\big)\\
\displaystyle \ +\sum_{n\in\bbN}\f{\big(\big(1+\nu+z^{1/2}\big)/2\big)_n\big(\big(1+\nu-z^{1/2}\big)/2\big)_n}{(n!)^2}\big(\sin^2(x)\big)^n \\
\ \times\big[\psi\big(n+\big(\big(1+\nu+z^{1/2}\big)/2\big)\big)-\psi\big(\big(1+\nu+z^{1/2}\big)/2\big) \\
\quad +\psi\big(n+\big(\big(1+\nu-z^{1/2}\big)/2\big)\big)-\psi\big(\big(1+\nu-z^{1/2}\big)/2\big)-2\psi(n+1)-2\gamma_E\big]\bigg\}, & \hspace{-.15cm} \mu=0,
\end{cases}
\end{split}\\
&\hspace{8cm} \nu\in[0,1),\; x\in(0,\pi/2),\; z\in\bbC.  \label{2.8}
\end{align}
Here, $\mathstrut_2F_1(\dott,\dott;\dott;\dott)$ denotes the hypergeometric function $($see, e.g., \cite[Ch.~15]{DLMF}$)$, $\psi(\dott) = \Gamma'(\dott)/\Gamma(\dott)$ denotes the digamma function, $\gamma_{E} = - \psi(1) = 0.57721\dots$ represents Euler's constant, and 
\begin{equation}
(\zeta)_0 =1, \quad (\zeta)_n = \Gamma(\zeta + n)/\Gamma(\zeta), \; n \in \bbN, 
\quad \zeta \in \bbC \backslash (-\bbN_0), 
\end{equation}
abbreviates Pochhammer's symbol $($see, e.g., \cite[Ch.~5]{DLMF}$)$. For more details on how to transform the Schr\"odinger equation $\tau_{\mu,\nu} u=zu$ with $\tau_{\mu,\nu}$ given in \eqref{2.1} to obtain these solutions from the hypergeometric differential equation, see Appendix \ref{sA}.

Notice that $\phi_{\frac{1}{2},\frac{1}{2}}(z,x,0)=z^{-1/2}\sin\big(z^{1/2}x\big)$ and $\theta_{\frac{1}{2},\frac{1}{2}}(z,x,0)=\cos\big(z^{1/2} x\big)$, in agreement with the fundamental set of solutions for the regular equation $\tau_{\frac{1}{2},\frac{1}{2}}u=zu$, while in general,
\begin{equation}\label{2.9}
    \theta_{\mu,\nu}(z,x,0)=(2\mu)^{-1}\phi_{-\mu,-\nu}(z,x,0),\quad \mu\in(0,1),\; \nu\in[0,1),\; x\in(0,\pi/2),\; z\in\bbC. 
\end{equation}

By defining the generalized boundary value operators for $f(z,\dott),f'(z,\dott)\in AC_{loc}((0,\pi/2))$ as,
\begin{align}\label{BCop}
\begin{split}
U_{0}(f)(z)&=\cos(\alpha)\wti f(z,0)+\sin(\alpha)\wti f^{\,\prime}(z,0),\\
U_{\pi/2}(f)(z)&=\cos(\beta)\wti f(z,\pi/2)-\sin(\beta)\wti f^{\,\prime}(z,\pi/2),
\end{split}
\end{align}
and 
\begin{align}\label{BCop1}
\begin{split}
V_{1}(f)(z)&=\wti f(z,0)-e^{i\varphi}R_{22}\wti f(z,\pi/2)+e^{i\varphi}R_{12}\wti f^{\,\prime}(z,\pi/2),\\
V_{2}(f)(z)&=\wti f^{\,\prime}(z,0)+e^{i\varphi}R_{21}\wti f(z,\pi/2)-e^{i\varphi}R_{11}\wti f^{\,\prime}(z,\pi/2),
\end{split}
\end{align}
the fundamental system of solutions introduced above is used to construct the characteristic function for the case of separated self-adjoint extensions as 
\begin{equation}\label{7}
F_{\mu,\nu}^{(\a,\b)}(z)=\cos(\a)\,U_{\pi/2}(\phi_{\mu,\nu})(z)-\sin(\a)\,U_{\pi/2}(\theta_{\mu,\nu})(z),
\end{equation}
while the one for coupled self-adjoint extensions can be found to be
\begin{equation}\label{9}
F_{\mu,\nu}^{(\varphi,R)}(z)=V_{1}(\theta_{\mu,\nu})(z)+V_{2}(\phi_{\mu,\nu})(z)+e^{2i\varphi}-1.
\end{equation}
 
The last step in evaluating $F_{\mu,\nu}^{(\a,\b)}(z)$ and $F_{\mu,\nu}^{(\varphi,R)}(z)$ consists, then, in finding the the generalized boundary values of $\phi_{\mu,\nu},\ \theta_{\mu,\nu}$ at the endpoint $x=\pi/2$. To achieve this, we need to utilize the linear transformations for the hypergeometric functions in order to extract the behavior of the fundamental system of solutions at $x=\pi/2$ from the one at the other endpoint $x=0$. For brevity, we will refer to these transformation as \textit{connection formulas} between the endpoints.
Using \cite[Eq. 15.8.4]{DLMF} yields the behavior near $x=\pi/2$ of the hypergeometric functions in \eqref{2.7} and \eqref{2.8} when $\nu\neq0$:
\begin{align}\lb{2.12}
\mathstrut_2F_1\big(&\big[1\pm(\mu+\nu)  + z^{1/2}\big]\big/2,\big[1\pm(\mu+\nu)  - z^{1/2}\big]\big/2;1\pm\mu;\sin^2 (x)\big) \no \\
&=\dfrac{\Gamma(1\pm\mu)\Gamma(\mp\nu)}{\Gamma\big(\big[1\pm(\mu-\nu)  + z^{1/2}\big]\big/2\big)\Gamma\big(\big[1\pm(\mu-\nu)  - z^{1/2}\big]\big/2\big)} \no \\
&\qquad \times \mathstrut_2F_1\big(\big[1\pm(\mu+\nu)  + z^{1/2}\big]\big/2,\big[1\pm(\mu+\nu)  - z^{1/2}\big]\big/2;1\pm\nu;\cos^2 (x)\big)  \\
&\quad\; + [\cos (x)]^{\mp2\nu}\dfrac{\Gamma(1\pm\mu)\Gamma(\pm\nu)}{\Gamma\big(\big[1\pm(\mu+\nu)  + z^{1/2}\big]\big/2\big)\Gamma\big(\big[1\pm(\mu+\nu)  - z^{1/2}\big]\big/2\big)} \no \\
&\qquad \times \mathstrut_2F_1\big(\big[1\pm(\mu-\nu)  + z^{1/2}\big]\big/2,\big[1\pm(\mu-\nu)  - z^{1/2}\big]\big/2;1\mp\nu;\cos^2 (x)\big), \no \\ 
&\hspace{4.5cm} \mu\in[0,1),\; \nu\in(0,1),\; x\in(0,\pi/2),\; z\in\bbC. \no 
\end{align}
Similarly, by \cite[Eq. 15.8.10]{DLMF}, one obtains for the case $\nu=0$ in \eqref{2.7}:
\begin{align}\label{2.13}
\mathstrut_2F_1\big(&\big[1\pm\mu  + z^{1/2}\big]\big/2,\big[1\pm\mu  - z^{1/2}\big]\big/2;1\pm\mu;\sin^2 (x)\big) \no\\
&=\frac{\Gamma(1\pm\mu)}{\Gamma\big(\big[1\pm\mu + z^{1/2}\big]\big/2\big)\Gamma\big(\big[1\pm\mu - z^{1/2}\big]\big/2\big)} 
\sum_{n=0}^\infty \dfrac{\big(\big[1\pm\mu + z^{1/2}\big]\big/2\big)_n\big(\big[1\pm\mu - z^{1/2}\big]\big/2\big)_n}{(n!)^2} 
  \\
&\quad\times \big[2\psi(n+1)  -\psi\big(n+\big[1\pm\mu + z^{1/2}\big]\big/2\big)-\psi\big(n+\big[1\pm\mu - z^{1/2}\big]\big/2\big) -\ln(\cos^2 (x))\big]
[\cos (x)]^{2n},   \no \\
& \hspace*{8.6cm}    \quad \mu\in[0,1),\; \nu=0,\; x\in(0,\pi/2),\; z\in\bbC. \no
\end{align}
The connection formula for the logarithmic solution $\theta_{0,\nu}$ is a bit more nuanced and it does not seem to have been given explicitly in the standard references. We direct the reader to Appendix \ref{sA1} for an outline of the derivation of the connection formula for this case.

By imposing on the fundamental set of solutions \eqref{2.7} and \eqref{2.8} the boundary values \eqref{2.4} and \eqref{2.5} and by using the connection formulas \eqref{2.12}, \eqref{2.13}, \eqref{A.16}, and \eqref{A.17} we obtain for $z\in\bbC$, the generalized values at $x=\pi/2$,
\begin{align}
\wti \phi_{\mu,\nu}(z,\tfrac{\pi}{2},0)&=\dfrac{2\Gamma(1+\mu)\Gamma(1+\nu)}{\Gamma\big(\big[1+\mu+\nu  + z^{1/2}\big]\big/2\big)\Gamma\big(\big[1+\mu+\nu  - z^{1/2}\big]\big/2\big)},\quad \mu,\nu\in[0,1),  \label{2.14} \\[3mm]
\wti \phi^{\, \prime}_{\mu,\nu}(z,\tfrac{\pi}{2},0)&=\begin{cases}
\dfrac{-\Gamma(1+\mu)\Gamma(-\nu)}{\Gamma\big(\big[1+\mu-\nu  + z^{1/2}\big]\big/2\big)\Gamma\big(\big[1+\mu-\nu  - z^{1/2}\big]\big/2\big)},&  \mu\in[0,1),\; \nu\in(0,1),\\[4mm]
\displaystyle\frac{\Gamma(1+\mu)\big[2\gamma_E  +\psi\big(\big[1+\mu + z^{1/2}\big]\big/2\big)+\psi\big(\big[1+\mu - z^{1/2}\big]\big/2\big)\big]}{\Gamma\big(\big[1+\mu + z^{1/2}\big]\big/2\big)\Gamma\big(\big[1+\mu - z^{1/2}\big]\big/2\big)}, & \mu\in[0,1),\; \nu=0,
\end{cases} \label{2.15} \\[3mm]
\wti \theta_{\mu,\nu}(z,\tfrac{\pi}{2},0)&=\begin{cases}
\dfrac{-\Gamma(-\mu)\Gamma(1+\nu)}{\Gamma\big(\big[1-\mu+\nu  + z^{1/2}\big]\big/2\big)\Gamma\big(\big[1-\mu+\nu  - z^{1/2}\big]\big/2\big)},&   \mu\in(0,1),\; \nu\in[0,1),\\[4mm]
\displaystyle\frac{\Gamma(1+\nu)\big[2\gamma_E  +\psi\big(\big[1+\nu + z^{1/2}\big]\big/2\big)+\psi\big(\big[1+\nu - z^{1/2}\big]\big/2\big)\big]}{\Gamma\big(\big[1+\nu + z^{1/2}\big]\big/2\big)\Gamma\big(\big[1+\nu - z^{1/2}\big]\big/2\big)}, & \mu=0,\; \nu\in[0,1) ,
\end{cases} \label{2.16} \\[3mm]
\wti \theta^{\, \prime}_{\mu,\nu}(z,\tfrac{\pi}{2},0)&=\begin{cases}
\dfrac{\Gamma(-\mu)\Gamma(-\nu)}{2\Gamma\big(\big[1-\mu-\nu  + z^{1/2}\big]\big/2\big)\Gamma\big(\big[1-\mu-\nu  - z^{1/2}\big]\big/2\big)},& \mu,\nu\in(0,1)\\[4mm]
\displaystyle\frac{-\Gamma(-\nu)\big[2\gamma_E  +\psi\big(\big[1-\nu + z^{1/2}\big]\big/2\big)+\psi\big(\big[1-\nu - z^{1/2}\big]\big/2\big)\big]}{2\Gamma\big(\big[1-\nu + z^{1/2}\big]\big/2\big)\Gamma\big(\big[1-\nu - z^{1/2}\big]\big/2\big)}, & \mu=0,\; \nu\in(0,1),\\[4mm]
\displaystyle\frac{-\Gamma(-\mu)\big[2\gamma_E  +\psi\big(\big[1-\mu + z^{1/2}\big]\big/2\big)+\psi\big(\big[1-\mu - z^{1/2}\big]\big/2\big)\big]}{2\Gamma\big(\big[1-\mu + z^{1/2}\big]\big/2\big)\Gamma\big(\big[1-\mu - z^{1/2}\big]\big/2\big)}, & \mu\in(0,1),\; \nu=0,\\[4mm]
(2\pi)^{-1} \cos\big(\pi z^{1/2}/2\big) \big(\big[2\gamma_E+\psi\big([1+z^{1/2}\big]/2\big)\\
\hspace{3cm} +\psi\big(\big[1-z^{1/2}\big]/2\big)\big]^2-\pi^2 \big[\cos\big(\pi z^{1/2}/2\big)\big]^{-2}
\big), & \mu=0=\nu.
\end{cases} \label{2.17}
\end{align}
The explicit expressions \eqref{2.14}-\eqref{2.17} used in the formulas \eqref{7} and \eqref{9} will allow us to obtain, through the integral representation \eqref{16}, the spectral $\zeta$-function for a given self-adjoint extension of $T_{\mu,\nu,min}$. 

We would like to make a comment about the self-adjoint extensions of the differential expression \eqref{2.1} when $\mu=\nu\neq0$.
In this case, \eqref{2.1} can be simplified to give
\begin{equation}
\tau_{\mu,\mu} = - \dfrac{d^2}{dx^2} +4\frac{\mu^2-(1/4)}{\sin^2(2x)},\quad  x\in(0,\pi/2),
\end{equation}
which, after a simple change of variables, represents a  Schr\"{o}dinger equation with a P\"oschl--Teller potential symmetric about the midpoint of the interval. Furthermore, the results in \eqref{2.15} and \eqref{2.16} show that for $\nu=\mu\in(0,1)$ and $z\in\bbC$, the generalized boundary values become
\begin{equation}\label{2.18}
\wti \phi^{\, \prime}_{\mu,\mu}(z,\tfrac{\pi}{2},0)=\wti \theta_{\mu,\mu}(z,\tfrac{\pi}{2},0)=\dfrac{-\Gamma(-\mu)\Gamma(1+\mu)}{\Gamma\big(\big[1  + z^{1/2}\big]\big/2\big)\Gamma\big(\big[1 - z^{1/2}\big]\big/2\big)}=[\sin(\pi\mu)]^{-1}\cos\big(\pi z^{1/2}/2\big),
\end{equation}
where the last equality follows from Euler's reflection formula (cf. \cite[no.~6.1.17]{AS72})
\begin{align}\lb{2.31}
\Gamma(w)\Gamma(1-w)=\dfrac{\pi}{\sin(\pi w)}, \quad w \in \bbC \backslash \bbZ. 
\end{align}
The result in \eqref{2.18} has interesting consequences for those self-adjoint extensions of $T_{\mu,\mu,min}$ whose characteristic function can be written as a linear combination of $\wti \phi^{\, \prime}_{\mu,\mu}(z,\tfrac{\pi}{2},0)$ and $\wti \theta_{\mu,\mu}(z,\tfrac{\pi}{2},0)$. For instance, in the separated case, the self-adjoint extensions $T_{\mu,\mu}^{(0,\pi/2)}$ and $T_{\mu,\mu}^{(\pi/2,0)}$ possess characteristic functions that simplify to 
\begin{equation}\lb{2.32}
    F_{\mu,\mu}^{(0,\pi/2)}(z)=-\wti \phi^{\, \prime}_{\mu,\mu}(z,\tfrac{\pi}{2},0),\quad
    F_{\mu,\mu}^{(\pi/2,0)}(z)=-\wti \theta_{\mu,\mu}(z,\tfrac{\pi}{2},0),\quad \mu\in(0,1).
\end{equation}
Moreover, in the coupled case, the self-adjoint extensions 
$T_{\mu,\mu}^{(\pi/2,\bar{R})}$, with $\bar{R}$ denoting the subgroup of $SL(2,\R)$ consisting of only diagonal matrices, are associated with characteristic functions of the form
\begin{equation}\lb{2.33}
F_{\mu,\mu}^{(\pi/2,\bar{R})}(z)=-i[R_{11}\wti \phi^{\, \prime}_{\mu,\mu}(z,\tfrac{\pi}{2},0)+R_{22}\wti \theta_{\mu,\mu}(z,\tfrac{\pi}{2},0)]=-i(R_{11}+R^{-1}_{11})\wti \phi^{\, \prime}_{\mu,\mu}(z,\tfrac{\pi}{2},0),\quad \mu\in(0,1).   
\end{equation}

The expressions \eqref{2.32} and \eqref{2.33} together with the result \eqref{2.18} show that the zeroes of $F_{\mu,\mu}^{(0,\pi/2)}(z)$, $F_{\mu,\mu}^{(\pi/2,0)}(z)$, and $F_{\mu,\mu}^{(\pi/2,\bar{R})}(z)$, and, consequently, the eigenvalues of the corresponding self-adjoint extensions, are \textit{independent of the strength $($that is the value of $\mu$$)$ of the potential}. In other words, the spectrum of $T_{\mu,\mu}^{(0,\pi/2)}$, $T_{\mu,\mu}^{(\pi/2,0)}$, and $T_{\mu,\mu}^{(\pi/2,\bar{R})}$ is not influenced by the presence of the potential. In fact, solving for the zeros of \eqref{2.18} in $z$ shows that the spectrum of the extensions $T_{\mu,\mu}^{(0,\pi/2)}$, $T_{\mu,\mu}^{(\pi/2,0)}$, and $T_{\mu,\mu}^{(\pi/2,\bar{R})}$ coincides with that of the Dirichlet-Neumann extension of the free problem, $\mu=1/2=\nu$ (i.e., zero potential case), on $(0,\pi/2)$. This further shows that the spectral $\zeta$-function for all of these cases is simply related to the Riemann $\zeta$-function via
\begin{equation}\label{2.34}
\zeta(s;T_{\mu,\mu}^{(0,\pi/2)})=\zeta(s;T_{\mu,\mu}^{(\pi/2,0)})=\zeta(s;T_{\mu,\mu}^{(\pi/2,\bar{R})})=\big(1-2^{-2s}\big)\zeta_R(2s),\quad \mu\in(0,1).
\end{equation}

Lastly, we would like to highlight the Friedrichs extension which is obtained by setting $\a=\b=0$ in \eqref{2.12text}. The corresponding characteristic function is easily found to be
\begin{equation}\label{2.22a}
    F_{\mu,\nu}^{(0,0)}(z)=\wti\phi_{\mu,\nu}(z,\tfrac{\pi}{2},0),\quad \mu,\nu\in[0,1),\; z\in\bbC.
\end{equation}
From the expression \eqref{2.14} it is easy to show that the zeroes of $F_{\mu,\nu}^{(0,0)}(z)$, and hence the eigenvalues of the Friedrichs extension, are explicitly given by
\begin{equation}\label{2.23a}
\sigma(T_{\mu,\nu}^{(0,0)})=\big\{[2n+1+\mu+\nu]^2\big\}_{n\in\bbN_0},\quad \mu,\nu\in[0,1).
\end{equation}
The spectral  $\zeta$-function associated with the Friedrichs extension is then the Hurwitz $\zeta$-function 
\begin{equation}\label{2.24}
\zeta(s;T_{\mu,\nu}^{(0,0)})=2^{-2s}\zeta_{H}\left(2s,\tfrac{1+\mu+\nu}{2}\right),\quad \mu,\nu\in[0,1).
\end{equation}
We remark that the Hurwitz $\zeta$-function is the spectral $\zeta$-function associated with any self-adjoint extension whose characteristic function can be expressed as a $z$-independent multiple of only one of the boundary values in \eqref{2.14}--\eqref{2.17}, excluding any parameter choices resulting in the presence of digamma functions. For instance, one readily verifies the following:
\begin{align}
\zeta(s;T_{\mu,\nu}^{(0,\pi/2)})&=2^{-2s}\zeta_{H}\left(2s,\tfrac{1+\mu-\nu}{2}\right),\quad \mu\in[0,1),\; \nu\in(0,1), \label{2.25}\\
\zeta(s;T_{\mu,\nu}^{(\pi/2,0)})&=2^{-2s}\zeta_{H}\left(2s,\tfrac{1-\mu+\nu}{2}\right),\quad \mu\in(0,1),\; \nu\in[0,1),\label{2.25a}\\
\zeta(s;T_{\mu,\nu}^{(\pi/2,\pi/2)})&=\begin{cases}2^{-2s}\zeta_{H}\left(2s,\frac{1-\mu-\nu}{2}\right),& \mu+\nu\neq 1,\\
2^{-2s}\zeta_{R}(2s),& \mu+\nu= 1,
\end{cases}\quad \mu,\nu\in(0,1).\label{2.25b}
\end{align}
This implies that in each of these cases, as well as those leading to \eqref{2.34}, the analytic continuation of the spectral $\zeta$-function to a meromorphic function in $s$ defined for all $s\neq 1/2$, is provided by the functional equations satisfied by $\zeta_R$ and $\zeta_H$ (see \cite[Eq. 25.4.1]{DLMF} and \cite[Eq. 25.11.9]{DLMF}, resp.). In particular, all of these spectral $\zeta$-functions will only have one simple pole at $s=1/2$ with residue given by the prefactor of each $\zeta_R$ and $\zeta_H$ times $1/2$. Furthermore, the $\zeta$-regularized determinant (see Section \ref{s4}) is well defined and can be easily evaluated by recalling that $\zeta_R(w)=\zeta_H(w,1)$ and by utilizing the formulas (see \cite[25.11.13]{DLMF} and \cite[Eq. 25.11.18]{DLMF}, resp.)
\begin{equation}\label{2.28}
\zeta_H(0,a)=(1/2)-a,\quad \tfrac{d}{dw}\zeta_H(w,a)\big|_{w=0}=\ln(\Gamma(a))-(1/2)\ln(2\pi),\quad \Re(a)>0.
\end{equation}
For instance, from \eqref{2.24} and \eqref{2.28} one readily verifies that
\begin{equation}\label{2.29}
\tfrac{d}{ds}\zeta(s;T_{\mu,\nu}^{(0,0)})|_{s=0}=(\mu+\nu)\ln 2+2\ln\left[\Gamma\left(\tfrac{1+\mu+\nu}{2}\right)\right]-\ln (2\pi),\quad \mu,\nu\in[0,1).
\end{equation}
This shows that the structure of the spectral $\zeta$-function in all of these cases does not differ from the case free of the potential.
In the next section, we will focus on analytically continuing the spectral $\zeta$-function for self-adjoint extensions that include the log solution, which will result in more exotic behaviors.

\section{The analytic continuation of the spectral \texorpdfstring{$\zeta$}{zeta}-function} \lb{s3}

The first step in the process of analytic continuation of $\zeta(s;T_{\mu,\nu}^{(A,B)})$ in \eqref{16} consists in shrinking the integration contour $\gamma$ along the branch cut $R_\Psi$ in \eqref{1a}. In the case of quasi-regular Sturm--Liouville problems, such as the the Schr\"odinger equation with P\"oschl--Teller potential considered here, it is possible to show that this process of contour deformation leads to the representation \cite[Sec. 3]{FPS24}
\begin{equation}\label{19a}
\zeta(s;T_{\mu,\nu}^{(A,B)})=e^{is(\pi-\Psi)}\frac{\sin(\pi s)}{\pi}\int_{0}^{\infty}dt\,t^{-s}\frac{d}{dt}\ln\left[(te^{i\Psi})^{-m_0}F_{\mu,\nu}^{(A,B)}\left(te^{i\Psi}\right)\right],
\end{equation} 
which is valid in the region $1/2<\Re(s)<1$. In order to obtain an expression of the spectral $\zeta$-function that is valid in a region to the left\footnote{An expression for the spectral $\zeta$-function which is valid for $\Re(s)>1/2$ is the one provided in \eqref{16}.} of $\Re(s)=1/2$ we need to subtract, and then add, from the integrand in \eqref{19a}, a number of terms of the large-$t$ asymptotic expansion of $\ln\big[(te^{i\Psi})^{-m_0}F_{\mu,\nu}^{(A,B)}\left(te^{i\Psi}\right)\big]$ (see, e.g., \cite{FGKS21,FGK,FPS24,Ki02,KM03,KM04}). By assuming that 
\begin{align}
    \ln\big[(te^{i\Psi})^{-m_0}F_{\mu,\nu}^{(A,B)}\left(te^{i\Psi}\right)\big]=\mathcal{L}^{(A,B)}_{asy,N}\left(te^{i\Psi},\mu,\nu\right)+O(t^{-N-1}),
\end{align}
which, according to Appendix \ref{sB}, is the general form of the asymptotic expansion of the cases of interest in this work, we can recast \eqref{19a} as
\begin{align}\lb{3.3ab}
\zeta(s;T_{\mu,\nu}^{(A,B)})&=e^{is(\pi-\Psi)}\frac{\sin(\pi s)}{\pi}\Bigg[\int_{0}^{1}dt\,t^{-s}\frac{d}{dt}\ln\left[(te^{i\Psi})^{-m_0}F_{\mu,\nu}^{(A,B)}\left(te^{i\Psi}\right)\right]\nonumber\\
&\quad+\int_{1}^{\infty}dt\,t^{-s}\frac{d}{dt}\left\{\ln\left[(te^{i\Psi})^{-m_0}F_{\mu,\nu}^{(A,B)}\left(te^{i\Psi}\right)\right]-\mathcal{L}^{(A,B)}_{asy,N}\left(te^{i\Psi},\mu,\nu\right)\right\} \Bigg]\nonumber\\
&\quad+ e^{is(\pi-\Psi)}\frac{\sin(\pi s)}{\pi}\int_{1}^{\infty}dt\,t^{-s}\frac{d}{dt}\mathcal{L}^{(A,B)}_{asy,N}\left(te^{i\Psi},\mu,\nu\right).
\end{align}
The first integral in \eqref{3.3ab} is convergent for $\Re(s)<1$ since $\ln[(te^{i\Psi})^{-m_0}F_{\mu,\nu}^{(A,B)}\left(te^{i\Psi}\right)]=O(t)$ as $t\to 0$ \cite[Sec. 3]{FPS24}. By using integration by parts, the second integral can be shown to be convergent for $\Re(s)>-(N+1)$, since the difference in curly parentheses is $O(t^{-N-1})$ at $t\to\infty$. The last integral in \eqref{3.3ab} is convergent for $\Re(s)>1/2$, since the characteristic function is of order $1/2$ (see, e.g., \cite[Prop. 3.5]{FPS24}), and can, in general, be computed explicitly as detailed later.
These remarks allow us to write the analytically continued expression of the spectral zeta function as
\begin{align}\label{3.4ab}
\zeta(s;T_{\mu,\nu}^{(A,B)})=\mathcal{F}_{\mu,\nu}^{(\a,\b)}(s,N)+e^{is(\pi-\Psi)}\frac{\sin(\pi s)}{\pi}\int_{1}^{\infty}dt\,t^{-s}\frac{d}{dt}\mathcal{L}^{(A,B)}_{asy,N}\left(te^{i\Psi},\mu,\nu\right), 
\end{align}
where $\mathcal{F}_{\mu,\nu}^{(\a,\b)}(s,N)$ is an entire function for $-(N+1)<\Re(s)<1$ defined by the sum of the first two integrals in \eqref{3.3ab}, while the second term of \eqref{3.4ab} encodes all the information regarding possible poles and branch points of the spectral $\z$-function.  

Since the main purpose of this paper is to show that the spectral $\z$-function can develop quite an exotic structure, we refrain from presenting the process of analytic continuation for all separated and coupled self-adjoint extensions and instead focus our analysis on particular cases that completely illustrate the atypical behavior of the spectral $\z$-function. The main conclusions of this section are summarized in Theorem \ref{thmstructure}.

We now consider the class of separated self-adjoint extensions identified by the choice of parameters $\mu=0$ and $\nu\in(0,1)$. 
By utilizing \eqref{2.14}-\eqref{2.17}, the characteristic function $F_{0,\nu}^{(\a,\b)}(z)$ associated with these self-adjoint extensions can be shown to be
\begin{align}\lb{3.2}
F_{0,\nu}^{(\a,\b)}(z)&=-\dfrac{2\Gamma(1+\nu)\cos(\b)}{\Gamma\big(\big[1+\nu  + z^{1/2}\big]\big/2\big)\Gamma\big(\big[1+\nu  - z^{1/2}\big]\big/2\big)}\nonumber\\
&\quad\times\left\{-\cos(\a)+\frac{\sin(\a)}{2}\big[2\gamma_E  +\psi\big(\big[1+\nu + z^{1/2}\big]\big/2\big)+\psi\big(\big[1+\nu - z^{1/2}\big]\big/2\big)\big]\right\}\nonumber\\
&\quad-\dfrac{\Gamma(-\nu)\sin(\b)}{\Gamma\big(\big[1-\nu  + z^{1/2}\big]\big/2\big)\Gamma\big(\big[1-\nu  - z^{1/2}\big]\big/2\big)}\nonumber\\
&\quad\times\left\{-\cos(\a)+\frac{\sin(\a)}{2}\big[2\gamma_E  +\psi\big(\big[1-\nu + z^{1/2}\big]\big/2\big)+\psi\big(\big[1-\nu - z^{1/2}\big]\big/2\big)\big]\right\}.
\end{align}

The existence of the zero eigenvalue in the self-adjoint extensions $T_{0,\nu}^{(\a,\b)}$, $\nu\in(0,1)$, and hence the value of $m_0$ in \eqref{19a}, can be determined by analyzing the small-$z$ expansion of its corresponding characteristic function. 
As $z\to 0$ one finds
\begin{align}\lb{3.5}
F_{0,\nu}^{(\a,\b)}(z)
&\underset{z\downarrow0}{=}-\dfrac{2\Gamma(1+\nu)\cos(\beta)}{\Gamma([1+\nu]/2)^2}\left\{-\cos(\a)+\sin(\a)[\gamma_E  +\psi([1+\nu]\big/2)]\right\} \\
&\quad\ \; -\dfrac{\Gamma(-\nu)\sin(\beta)}{\Gamma([1-\nu]/2)^2}\left\{-\cos(\a)+\sin(\a)[\gamma_E  +\psi([1-\nu]\big/2)]\right\}+O(z),\quad \nu\in(0,1),\; \al,\b\in[0,\pi).\no
\end{align}
Therefore, given $\nu\in(0,1)$, $z=0$ is a zero of $F_{0,\nu}^{(\a,\b)}(z)$ if and only if, for some $\alpha,\beta\in(0,\pi)$ the constant term in \eqref{3.5} is zero.

In order to proceed, we now restrict the analysis to the setting $\nu=p/q$ with $p,q\in\N$, $0< p\leq q-1$, and $p,q$ relatively prime. This constraint is needed to perform explicitly the analytic continuation of the spectral $\zeta$-function. We would like to add that one can follow an argument similar to the one leading to the asymptotic expansion \eqref{B45a} (necessary for the analytic continuation) for a fixed value of $\nu$ that lies outside of this class (see Remark \ref{RemB} for more details). The large-$z$ asymptotic expansion of the logarithm of the characteristic function \eqref{3.2}, needed for the analytic continuation of the associated spectral $\z$-function, are obtained from the results presented in Section \ref{SubB3} of Appendix \ref{sB}. 
We begin with the case $\b\neq0$.

\subsection{Case \texorpdfstring{$\b\neq0$}{notzero}} 
By setting $z=te^{i\Psi}$ in \eqref{B45b} and by differentiating the ensuing expression with respect to $t$,  we obtain, for the case $\b\neq 0$, the expansion (with $q_N=\textrm{min}\{q,\lceil q(N+1)/p\rceil-1\}$)
\begin{align}\lb{3.5a}
\frac{d}{dt}\mathcal{L}^{(\a,\beta)}_{asy,N}(te^{i\Psi},0,p/q)&=-\frac{i\pi e^{i\Psi/2}}{4}t^{-1/2}-\sum_{n=0}^{N}\mathcal{B}_{n}(p/q,t)t^{-n-1}e^{-i n\Psi}\nonumber\\
&\quad+\sum_{m=1}^{q_N}\frac{(-1)^{m}p}{q}[\Omega_{0}\left(p/q\right)]^{m}t^{-\frac{mp}{q}-1}e^{-i \frac{mp}{q}\Psi}\nonumber\\
&\quad-\sum_{\underset{m\neq jp\,,0\leq j\leq q_N-1}{m=q}}^{q(N+1)-p-1}\mathcal{C}_{m}(p/q,\b,t)t^{-\frac{m+p}{q}-1}e^{-i \frac{m+p}{q}\Psi},   
\end{align}
where, by introducing $\eta_\a=[\g_E-\ln 2-(i/2)(\pi-\Psi)]\sin(\a)-\cos(\a)$,  we find
\begin{align}
 \mathcal{B}_{0}(p/q,t)=-\left(\frac{p}{2q}-m_0\right)- \frac{\sin(\a)}{2[\eta_\a+\sin(\a)\ln(t^{1/2})]},  
\end{align}
and by differentiating $\mathcal{R}_{m}(p/q,te^{i\Psi})(te^{i\Psi})^{-m}$ with respect to $t$ and using \eqref{B.36} we obtain, for $n\geq 1$,
\begin{align}
\mathcal{B}_{n}(p/q,t)&=n p_{n,0}(p/q)\\
&\quad +\sum_{j=1}^{n}p_{n,j}(p/q)(\sin(\a))^{j}\Big[n(\eta_\a+\sin(\a)\ln(t^{1/2}))^{-j}+\frac{j\sin(\a)}{2}(\eta_\a+\sin(\a)\ln(t^{1/2}))^{-j-1}\Big].\notag
\end{align}
Similarly, by differentiating $\mathcal{S}_{m}(p/q,z)(te^{i\Psi})^{-(m+p)/q}$ with respect to $t$ and by taking into account \eqref{B45}, we find
\begin{align}
\mathcal{C}_{m}(p/q&,\b,t)=\frac{m+p}{q}g_{m,0}(p/q,\b)\\
&+ \sum_{j=1}^{k_m}g_{m,j}(p/q,\b)(\sin(\a))^{j}\Big[\frac{m+p}{q}(\eta_\a+\sin(\a)\ln(t^{1/2}))^{-j}+\frac{j\sin(\a)}{2}(\eta_\a+\sin(\a)\ln(t^{1/2}))^{-j-1}\Big].\nonumber
\end{align}
By using \eqref{3.5a} in the expression \eqref{3.4ab} we obtain the following result 
(cf. \cite[Sec. 3.3]{FGKS21} and \cite[Sec. 4]{FPS24}):
\begin{align}\lb{3.9a}
    \zeta(s;T_{0,p/q}^{(\a,\b)})=\mathcal{F}_{0,p/q}^{(\a,\b)}(s,N)+e^{is(\pi-\Psi)}\frac{\sin(\pi s)}{\pi}\Bigg[-i\frac{(\pi/2) e^{i\Psi/2}}{2s-1}+\sum_{m=1}^{q_N}\frac{(-1)^{m}p}{q}[\Omega_{0}\left(p/q\right)]^{m}\frac{e^{i\frac{mp}{q}\Psi}}{s+\frac{mp}{q}}\nonumber\\
    -\sum_{n=0}^{N}e^{-in\Psi}\int_{1}^{\infty}dt\,t^{-s-n-1}\mathcal{B}_{n}(p/q,t)-\sum_{\underset{m\neq jp\,,0\leq j\leq q_N-1}{m=q}}^{q(N+1)-p-1}e^{-i\left(\frac{m+p}{q}\right)\Psi}\int_{1}^{\infty}dt\,t^{-s-\frac{m+p}{q}-1}\mathcal{C}_{m}(p/q,\b,t)\Bigg],
\end{align}
where the function 
\begin{align}\lb{3.10}
\mathcal{F}_{0,p/q}^{(\a,\b)}(s,N)&=e^{is(\pi-\Psi)}\frac{\sin(\pi s)}{\pi}\int_{0}^{1}dt\,t^{-s}\frac{d}{dt}\ln\left[(te^{i\Psi})^{-m_0}F_{0,p/q}^{(\a,\b)}\left(te^{i\Psi}\right)\right]\\
&\quad+e^{is(\pi-\Psi)}\frac{\sin(\pi s)}{\pi}\int_{1}^{\infty}dt\,t^{-s}\frac{d}{dt}\Bigg\{\ln\left[(te^{i\Psi})^{-m_0}F_{0,p/q}^{(\a,\b)}\left(te^{i\Psi}\right)\right]-\mathcal{L}^{(\a,\beta)}_{asy,N}(te^{i\Psi},0,p/q)\Bigg\},   \nonumber 
\end{align}
is entire for $-(N+1)<\Re(s)<1$.
The integrals involving $\mathcal{B}_{n}(p/q,t)$ and $\mathcal{C}_{m}(p/q,\b,t)$ on the right-hand side of \eqref{3.9a} can be computed according to the formula 
\begin{equation}\label{3.11a}
\int_{1}^{\infty}dt\,t^{-s-n-1}\left(\eta_{\a}+\sin(\a)\ln\,t^{1/2}\right)^{-k}=\frac{2\eta_{\a}^{1-k}}{\sin(\a)}e^{\frac{2(s+n)\eta_{\a}}{\sin(\a)}}E_{k}\left(\frac{2(s+n)\eta_{\a}}{\sin(\a)}\right),\quad n,k\in\N,
\end{equation}
which can be obtained by using the definition of the 
generalized exponential integral function $E_{k}(z)$ in \cite[Eq. 8.19.3]{DLMF} after performing the change of variables $x=\eta_{\a}+\sin(\a)\ln\,t^{1/2}$.

We point out that the choice $\alpha=0$ is valid through \eqref{3.10} and on the left-hand side of \eqref{3.11a}. However, on the right-hand side of \eqref{3.11a}, one must consider the $\alpha\downarrow 0$ limit. This can be accomplished by employing \cite[Eqs. 8.19.16 and 13.2.6]{DLMF} to write
\begin{equation}\label{3.12a}
e^{\frac{2(s+n)\eta_{\a}}{\sin(\a)}}E_{k}\left(\frac{2(s+n)\eta_{\a}}{\sin(\a)}\right)\underset{\a\downarrow0}{=}\frac{\sin(\a)}{2(s+n)\eta_{\a}}.
\end{equation}
Hence, the choice $\a=0$ in what follows is understood in this limiting sense.

By utilizing \eqref{3.11a} we find the expression for the spectral $\z$-function valid for $-(N+1)<\Re(s)<1$
\begin{align}\label{3.12}
\zeta(s;T_{0,p/q}^{(\a,\b)})&=\cF_{0,p/q}^{(\a,\b)}(s,N)+e^{is(\pi-\Psi)}\frac{\sin(\pi s)}{\pi}\Bigg\{-i\frac{(\pi/2) e^{i\Psi/2}}{2s-1}+\sum_{m=1}^{q_N}\frac{(-1)^{m}p}{q}[\Omega_{0}\left(p/q\right)]^{m}\frac{e^{i\frac{mp}{q}\Psi}}{s+\frac{mp}{q}}\nonumber\\
&\quad+\left(\frac{p/q}{2}-m_0\right)s^{-1}+e^{\frac{2s\lambda}{\sin(\a)}}E_{1}\left(\frac{2s\lambda}{\sin(\a)}\right)-\sum_{n=1}^{N}e^{-i n \Psi}\Bigg[\frac{n p_{n,0}(p/q)}{s+n}+\Xi(s,p/q,n)\Bigg]\nonumber\\
&\quad-\sum_{\underset{m\neq jp\,,0\leq j\leq q_N-1}{m=q}}^{q(N+1)-p-1}e^{-i \frac{m+p}{q}\Psi}\Bigg[\frac{m+p}{q}\frac{ g_{m,0}(p/q,\b)}{s+\frac{m+p}{q}}+\Delta\left(s,p/q,\frac{m+p}{q}\right)\Bigg]\Bigg\},
\end{align}
where we have defined, for convenience, the functions 
\begin{align}\lb{3.13}
 \Xi(s,p/q,n)&= \sum_{k=1}^{n}\frac{2p_{n,k}(p/q)(\sin(\a))^{k-1}}{\eta_{a}^{k}}e^{\frac{2(s+n)\eta_{a}}{\sin(\a)}} \\
&\quad \,\times\left[\frac{k\sin(\a)}{2}E_{k+1}\left(\frac{2(s+n)\eta_{a}}{\sin(\a)}\right)+n\eta_{a} E_{k}\left(\frac{2(s+n)\eta_{a}}{\sin(\a)}\right)\right], \notag
\end{align}
and
\begin{align}\lb{3.14}
   \Delta\left(s,p/q,\frac{m+p}{q}\right)&=\sum_{k=1}^{k_m}\frac{2g_{m,k}(p/q,\b)(\sin(\a))^{k-1}}{\eta_{a}^{k}}e^{\frac{2\left(s+\frac{m+p}{q}\right)\eta_{a}}{\sin(\a)}}\\
   &\quad\times
\left[\frac{k\sin(\a)}{2}E_{k+1}\left(\frac{2\left(s+\frac{m+p}{q}\right)\eta_{a}}{\sin(\a)}\right)+\frac{m+p}{q}\eta_{a} E_{k}\left(\frac{2\left(s+\frac{m+p}{q}\right)\eta_{a}}{\sin(\a)}\right)\right].\notag
\end{align}

The result \eqref{3.12} shows explicitly the very unusual structure of the spectral $\z$-function associated with the self-adjoint extensions $T_{0,p/q}^{(\a,\b)}$. We point out first that a simple pole appears for all self-adjoint extensions at the point $s=1/2$, as it is to be expected according to \cite[Lemma 4.4]{FPS24}.

We must now distinguish the following two exceptional cases: $\a=0$ and $\b=\pi/2$. Before proceeding, we note that taking $N$ large enough in \eqref{3.12} will give $q_N=q$ and the other two sums in \eqref{3.12} will simply include more of the structure as $N$ grows larger. This then illustrates the general structure of the spectral $\zeta$-function, which we now analyze by assuming that the continuation is done to all orders (i.e., as $N\to\infty$).

\subsubsection{\texorpdfstring{Case $\a\neq0,\ \b\neq\pi/2$}{zero}}\label{sub3.1.1}  Suppose $\a\in(0,\pi)$ and $\b\in(0,\pi)\backslash\{\pi/2\}$. Then, in addition to $s=1/2$, simple poles occur also at the points $s=-mp/q$ with $1\leq m< q$ from the first line of \eqref{3.12} (since $\Omega_0(\nu)\neq0$ for $\nu\in(0,1)$ by \eqref{B26} for $\b\in(0,\pi)\backslash\{\pi/2\}$).

To analyze the behavior of the remaining non-trivial functions in \eqref{3.12}, note that \cite[Eq. 8.19.8]{DLMF}
\begin{equation}\lb{3.15}
E_{k}\left(\frac{2(s+j)\eta_{\a}}{\sin(\a)}\right)=\frac{(-1)^{k}}{(k-1)!}\left(\frac{2(s+j)\eta_{\a}}{\sin(\a)}\right)^{k-1}\ln\left(\frac{2(s+j)\eta_{\a}}{\sin(\a)}\right)+O(1).
\end{equation}     
The result \eqref{3.15} along with \eqref{3.13} implies that the spectral $\z$-function \eqref{3.12} develops logarithmic branch points at $s=-j$ with $j\in\bbN_0$, where we have used the fact that at least $p_{n,n}(p/q)\neq0$ in \eqref{3.13} (see  Remark \ref{RemB2}).

To study the rest of the structure, one needs to obtain information about the functions $g_{m,k}(p/q,\beta)$ in \eqref{3.14}. If at least one of the functions $g_{m,0}(p/q,\beta)\neq0$ while the others indexed with $k\geq 1$ vanish identically, then the spectral $\z$-function develops a simple pole at $s=-(m+p)/q$. If, on the other hand, at least one of the functions $g_{m,k}(p/q,\beta)\neq0$ with $k\geq 1$, then a branch point appears at the point $s=-(m+p)/q$. Unfortunately, the functions $g_{m,k}(p/q,\beta)$ cannot be obtained in general (see Remark \ref{RemB3}) so we cannot state whether the points $s=-(m+p)/q$ correspond to simple poles or branch points. However, the analysis of the poles and branch points can be performed completely once $p,\ q,$ and $\beta$ are given. 

We now suppose that $g_{m,0}(p/q,\b)\neq0$ in \eqref{3.12} and at least one of $g_{m,k}(p/q,\beta)\neq0$ in \eqref{3.14}, which we will call the \textit{generic} case.\footnote{In principle, there could also be cancellations when branch points (or poles) from different terms of the expansion \eqref{3.12} coincide, such as when $-(M+p)/q=-j,\ j\in\bbN$. Since such cancellations have not been observed for particular values tested, we also exclude these when considering the generic case.} In the generic case, branch points also occur at $s=-(M+p)/q$ with $M\geq q$ and $M\neq jp$ for $0\leq j\leq q-1$. We point out that these branch points and the previously found poles never coincide since that would imply $M=p(m-1)$, but $1\leq m\leq q$ and $M\neq jp$ for $0\leq j\leq q-1$.

We would like to emphasize that the presence of logarithmic branch points at more than one point is an exotic and interesting feature associated with the spectral $\z$-function. Examples of similar behavior of the spectral $\z$-function are exceedingly scarce in the literature. In fact, this behavior has been found (with branch points only at nonpositive integers) for a class of self-adjoint extensions of the Legendre operator \cite[Sec. 5.2]{FPS24} and for the Laplace operator on higher-dimensional conical manifolds \cite{Ki08}.

\subsubsection{\texorpdfstring{Case $\a=0,\ \b\neq\pi/2$}{zero}}\label{sub3.1.2} Next, we consider the case $\a=0$ and $\b\in(0,\pi)\backslash\{\pi/2\}$. We point out, once again, that the choice $\a=0$ is understood as taking the limit $\a\downarrow0$ and employing \eqref{3.12a}. In particular, \eqref{3.12a} shows that the exponential integral term in \eqref{3.12}, as well as all of \eqref{3.13} and \eqref{3.14}, become zero when letting $\a\downarrow0$, 
so there are no branch points when $\a=0$ (as expected since this is equivalent to choosing the Friedrichs extension boundary condition at $x=0$). 

In addition to the omnipresent simple pole at $s=1/2$, the particular structure of the spectral $\z$-function depends on whether $g_{m,0}(p/q,\b)\neq0$ in \eqref{3.12}. Thus, in the generic case (see above), there are additional simple poles at $s=-mp/q$ with $1\leq m< q$ and at $s=-(M+p)/q$ with $M\geq q$, $M\neq jp$ for $0\leq j\leq q-1$, and $M\neq kq-p$ for $k\in \bbN$. The last constraint guarantees that none of the points $s=-(M+p)/q$ is a negative integer since, at those points, the spectral $\z$-function is regular thanks to the presence, in \eqref{3.12}, of the overall $\sin(\pi s)$ factor.

\subsubsection{\texorpdfstring{Case $\a\neq0,\ \b=\pi/2$}{zero}} For the other exceptional case, namely $\b=\pi/2$ and $\a\in(0,\pi)$, notice that by \eqref{B26}, $\Omega_0(\nu)=0=\Omega_k(\nu,z)$, for $k\geq 1$, so the sum involving the function $\Omega_0$ in the first line of \eqref{3.12} becomes zero. Moreover, \eqref{B33} shows that the functions $S_m$ vanish identically when $\b=\pi/2$, thus we must have that $g_{n,j}(\nu,\pi/2)\equiv 0$ by \eqref{B45}. Hence, in this case, the last line of \eqref{3.12} is also zero, so there is only one simple pole at $s=1/2$ along with logarithmic branch points at $s=-j$ with $j\in\N_0$.
One readily verifies that this structure actually holds for all $\nu\in(0,1)$ without the restriction to rationals since the terms that needed this restriction in Appendix \ref{sB} are exactly the $\Omega_k$ terms which do not occur when $\b=\pi/2$.

\subsubsection{\texorpdfstring{Case $\a=0,\ \b=\pi/2$}{zero}} Finally, taking $\a\downarrow0$ and $\beta=\pi/2$ in \eqref{3.12} makes all of the terms that account for the additional pole and branch point structure disappear, resulting in an expansion for all $\nu\in(0,1)$ which agrees with that of the Hurwitz $\zeta$-function given in \eqref{2.25}.

\subsection{\texorpdfstring{Case $\b=0$}{zero}} 
In this case, the terms needed for the analytic continuation can be obtained from \eqref{B51}. In particular, one has
\begin{align}\lb{3.16a}
\frac{d}{dt}\mathcal{L}^{(\a,0)}_{asy,N}(te^{i\Psi},0,\nu)&=-\frac{i\pi e^{i\Psi/2}}{4}t^{-1/2}-\sum_{n=0}^{N}\mathcal{B}_{n}(-\nu,t)t^{-n-1}e^{-i n\Psi},\quad \nu\in(0,1),\ \alpha\in(0,\pi).   
\end{align}
Similarly to the case $\beta\neq 0$, by utilizing \eqref{3.16a} in \eqref{3.4ab} we obtain the analytically continued expression for the spectral $\z$-function valid for $-(N+1)<\Re(s)<1$ 
\begin{align}\lb{3.17}
    \zeta(s;T_{0,\nu}^{(\a,0)})&=\mathcal{F}_{0,\nu}^{(\a,0)}(s,N)+e^{is(\pi-\Psi)}\frac{\sin(\pi s)}{\pi}\Bigg\{-i\frac{(\pi/2) e^{i\Psi/2}}{2s-1}
    -\left(\frac{\nu}{2}+m_0\right)s^{-1}+e^{\frac{2s\eta_\a}{\sin(\a)}}E_{1}\left(\frac{2s\eta_\a}{\sin(\a)}\right)\nonumber\\
    &\quad-\sum_{n=1}^{N}e^{-i n \Psi}\Bigg[\frac{n p_{n,0}(-\nu)}{s+n}+\Xi(s,-\nu,n)\Bigg]\Bigg\},
\end{align}
with 
\begin{align}\lb{3.18}
    \mathcal{F}_{0,\nu}^{(\a,0)}(s,N)&=e^{is(\pi-\Psi)}\frac{\sin(\pi s)}{\pi}\int_{0}^{1}dt\,t^{-s}\frac{d}{dt}\ln\left[(te^{i\Psi})^{-m_0}F_{0,\nu}^{(\a,0)}\left(te^{i\Psi}\right)\right]\\
&\quad+e^{is(\pi-\Psi)}\frac{\sin(\pi s)}{\pi}\int_{1}^{\infty}dt\,t^{-s}\frac{d}{dt}\Bigg\{\ln\left[(te^{i\Psi})^{-m_0}F_{0,\nu}^{(\a,0)}\left(te^{i\Psi}\right)\right]-\mathcal{L}^{(\a,0)}_{asy,N}(te^{i\Psi},0,\nu)\Bigg\},   \nonumber 
\end{align}
entire for $-(N+1)<\Re(s)<1$ and $\mathcal{L}^{(\a,0)}_{asy,N}(te^{i\Psi},0,\nu)$ given in \eqref{B51} once $z=te^{i\Psi}$. One sees from \eqref{3.17} that, in addition to the simple pole at $s=1/2$, the spectral $\z$-function develops logarithmic branch points at non-positive integers also in the case $\b=0$, except when $\a=0$ (which is, once again, understood as $\a\downarrow0$ by employing \eqref{3.12a}). In fact, taking $\a\downarrow0$ in \eqref{3.17} leads to an asymptotic expansion that agrees with that of the Hurwitz $\zeta$-function given in \eqref{2.24}.

We point out that this analysis in fact shows that the only spectral $\zeta$-functions exhibiting the ``usual" structure in the above cases are exactly those found in Section \ref{s2}, namely those in \eqref{2.24} and \eqref{2.25}.

\section{The derivative of the spectral \texorpdfstring{$\zeta$}{zeta}-function at zero} \lb{s4}

In this section we discuss the $\zeta$-regularized functional determinant for the parameter choices and self-adjoint extensions considered in Section \ref{s3}. Since we are focusing on the atypical behaviors, we begin with the case $\alpha\in(0,\pi)$ since, according to our previous analysis, $s=0$ is always a regular point when $\a=0$. The results of this section are summarized in Theorem \ref{thmdet}.

In order to evaluate the regularized functional determinant of the self-adjoint extension $T^{(\a,\b)}_{0,\nu}$ we need to extend the analytic continuation of its spectral $\z$-function to a region of the complex plane that contains the point $s=0$. This is accomplished, as we have seen in the previous section, by subtracting, and then adding, to the integrand in the representation \eqref{19a} a suitable number of terms of the large-$t$ asymptotic expansion of the logarithm of the characteristic function $(te^{i\Psi})^{-m_0}F_{0,\nu}^{(\a,\b)}(te^{i\Psi})$. Since we need the analytic continuation in a neighborhood of $s=0$ it is sufficient to focus only on the terms of the large-$t$ asymptotic expansion up to and including those of order $t^{-1}$. For this reason we do not need the restriction that $\nu$ be a rational number, and we can treat the case $\nu\in(0,1)$. Taking these comments into account, we have from \eqref{B45b}, with $p/q$ replaced by $\nu$ and $z=t e^{i\Psi}$, the expression
\begin{align}\lb{4.1}
\frac{d}{dt}\mathcal{L}^{(\a,\b)}_{asy,\nu}(te^{i\Psi},0,\nu)&=-\frac{i\pi e^{i\Psi/2}}{4}t^{-1/2}-\mathcal{B}_{0}(\nu,t)t^{-1},\quad \nu\in(0,1),\ \a,\b\in(0,\pi),
\end{align}
which yields, by the methods of Section \ref{s3}, the following analytic continuation valid for $-\nu<\Re(s)<1$:
\begin{align}\lb{4.2}
    \zeta(s;T_{0,\nu}^{(\a,\b)})=\mathcal{F}_{0,\nu}^{(\a,\b)}(s)+e^{is(\pi-\Psi)}\frac{\sin(\pi s)}{\pi}\Bigg[-i\frac{(\pi/2) e^{i\Psi}}{2s-1}
    +\left(\frac{\nu}{2}-m_0\right)s^{-1}+e^{\frac{2s\eta_\a}{\sin(\a)}}E_{1}\left(\frac{2s\eta_\a}{\sin(\a)}\right)\Bigg],
\end{align}
with 
\begin{align}\lb{4.3}
\mathcal{F}_{0,\nu}^{(\a,\b)}(s)&=e^{is(\pi-\Psi)}\frac{\sin(\pi s)}{\pi}\int_{0}^{1}dt\,t^{-s}\frac{d}{dt}\ln\left[(te^{i\Psi})^{-m_0}F_{0,\nu}^{(\a,\b)}\left(te^{i\Psi}\right)\right]\\
&\quad + e^{is(\pi-\Psi)}\frac{\sin(\pi s)}{\pi}\int_{1}^{\infty}dt\,t^{-s}\frac{d}{dt}\Bigg\{\ln\left[(te^{i\Psi})^{-m_0}F_{0,\nu}^{(\a,\b)}\left(te^{i\Psi}\right)\right]-\mathcal{L}^{(\a,\b)}_{asy,\nu}(te^{i\Psi},0,\nu)\Bigg\}, \no
\end{align}
being an entire function for $-\nu<\Re(s)<1$. The expression \eqref{4.2} can be utilized to obtain, according to \eqref{modzeta}, the regularized spectral $\z$-function. In fact, by noting that 
\begin{align}\lb{4.4}
  e^{is(\pi-\Psi)}\frac{\sin(\pi s)}{\pi}\left[e^{\frac{2s\eta_\a}{\sin(\a)}}E_{1}\left(\frac{2s\eta_\a}{\sin(\a)}\right)\right] 
  =-s\,\ln s-s\left[\ln\left(\frac{2\eta_\a}{\sin(\a)}\right)+\gamma_E\right]+O(s^{2}\ln s),
\end{align}
we obtain
\begin{align}\lb{4.5}
  \zeta_{\textrm{reg}}(s;T_{0,\nu}^{(\a,\b)})&= \mathcal{F}_{0,\nu}^{(\a,\b)}(s)+e^{is(\pi-\Psi)}\frac{\sin(\pi s)}{\pi}\left[-i\frac{(\pi/2) e^{i\Psi}}{2s-1}
    +\left(\frac{\nu}{2}-m_0\right)s^{-1}\right]\nonumber\\
   &\quad- s\left[\ln\left(\frac{2\eta_\a}{\sin(\a)}\right)+\gamma_E\right]+O(s^{2}\ln s).
\end{align}
The derivative of the function $\mathcal{F}_{0,\nu}^{(\a,\b)}(s)$ can be computed as follows 
\begin{align}\lb{4.6}
    \frac{d}{ds}\left[\mathcal{F}_{0,\nu}^{(\a,\b)}(s)\right]\bigg|_{s=0}&= \int_{0}^{1}dt\,\frac{d}{dt}\ln\left[(te^{i\Psi})^{-m_0}F_{0,\nu}^{(\a,\b)}\left(te^{i\Psi}\right)\right]\nonumber\\
&\quad+ \int_{1}^{\infty}dt\,\frac{d}{dt}\Bigg\{\ln\left[(te^{i\Psi})^{-m_0}F_{0,\nu}^{(\a,\b)}\left(te^{i\Psi}\right)\right]+\frac{i\pi e^{i\Psi/2}}{2}t^{1/2}\nonumber\\
&\quad-\ln\left[-\frac{\Gamma(-\nu)\sin(\b)e^{-i\pi\nu/2}}{2^{\nu+1}\pi}\right] -\left(\frac{\nu}{2}-m_0\right) \ln t-\left(\frac{\nu}{2}-m_0\right) i\Psi\nonumber\\
&\quad-\ln\left[\eta_\a+\sin(\a)\ln t^{1/2}\right]\Bigg\}\\
&=-\ln\big(\mathcal{F}^{(\a,\b)}_{0,\nu; m_0}\big)-\frac{i\pi e^{i\Psi/2}}{2}+ \ln\left[-\frac{\Gamma(-\nu)\sin(\b)e^{-i\pi\nu/2}}{2^{\nu+1}\pi}\right] \no \\
&\quad +\left(\frac{\nu}{2}-m_0\right) i\Psi +\ln\eta_\a, \no
\end{align}
where we have defined 
\begin{align}
 \lim_{t\downarrow 0}\,(te^{i\Psi})^{-m_0}F_{0,\nu}^{(\a,\b)}\left(te^{i\Psi}\right)=\mathcal{F}^{(\a,\b)}_{0,\nu; m_0}\neq 0.  
\end{align}

By using \eqref{4.6} and by differentiating the remaining terms in \eqref{4.5} with respect to the variable $s$ we obtain, for $\a,\b\in(0,\pi)$ and $\nu\in(0,1)$
\begin{align}\lb{4.9a}
   \tfrac{d}{ds}\zeta_{\textrm{reg}}(s;T_{0,\nu}^{(\a,\b)})|_{s=0} =-\ln\big(\mathcal{F}^{(\a,\b)}_{0,\nu; m_0}\big)+\ln\left[-\frac{\Gamma(-\nu)\sin(\a)\sin(\b)}{2^{\nu+2}\pi}\right]-i\pi m_0-\gamma_E. 
   \end{align}

For $\b=0$ the regularized spectral $\z$-function can be written, in a neighborhood of $s=0$, according to \eqref{3.17} and \eqref{3.18}, as
\begin{align}\lb{4.10}
  \zeta_{\textrm{reg}}(s;T_{0,\nu}^{(\a,0)})&= \mathcal{F}_{0,\nu}^{(\a,0)}(s)+e^{is(\pi-\Psi)}\frac{\sin(\pi s)}{\pi}\left[-i\frac{(\pi/2) e^{i\Psi}}{2s-1}
    -\left(\frac{\nu}{2}+m_0\right)s^{-1}\right]\nonumber\\
   &\quad- s\left[\ln\left(\frac{2\eta_\a}{\sin(\a)}\right)+\gamma_E\right]+O(s^{2}\ln s),
\end{align}
where, by using \eqref{B51} and an argument similar to the one leading to \eqref{4.6}, one finds
\begin{align}\lb{4.11}
\frac{d}{ds}\left[\mathcal{F}_{0,\nu}^{(\a,0)}(s)\right]\bigg|_{s=0}&=-\ln\big(\mathcal{F}^{(\a,0)}_{0,\nu; m_0}\big)-\frac{i\pi e^{i\Psi/2}}{2}+ \ln\left[-\frac{\Gamma(1+\nu)2^{\nu}e^{i\pi\nu/2}}{\pi}\right] -\left(\frac{\nu}{2}+m_0\right) i\Psi +\ln\eta_\a.
\end{align}
The results in \eqref{4.10} and \eqref{4.11} allow us to conclude that when $\b=0$, $\a\in(0,\pi)$, and $\nu\in(0,1)$,
\begin{align}\lb{4.12}
   \tfrac{d}{ds}\zeta_{\textrm{reg}}(s;T_{0,\nu}^{(\a,0)})|_{s=0} =-\ln\big(\mathcal{F}^{(\a,0)}_{0,\nu; m_0}\big)+\ln\left[-\frac{\Gamma(1+\nu)\sin(\a)2^{\nu-1}}{\pi}\right]-i\pi m_0-\gamma_E. 
   \end{align}

Lastly, for $\a=0$ and $\b\in(0,\pi)$, the spectral $\z$-function is regular at $s=0$ and its derivative can be found to be
\begin{align}\lb{4.13}
     \tfrac{d}{ds}\zeta(s;T_{0,\nu}^{(0,\b)})|_{s=0} =-\ln\big(\mathcal{F}^{(0,\b)}_{0,\nu; m_0}\big)+\ln\left[\frac{\Gamma(-\nu)\sin(\b)}{2^{\nu+1}\pi}\right]-i\pi m_0. 
\end{align}
We would like to point out that, when $\b=\pi/2$, \eqref{4.13} reduces to 
\begin{align}\lb{4.14}
     \tfrac{d}{ds}\zeta(s;T_{0,\nu}^{(0,\pi/2)})|_{s=0} =2\ln\left[\Gamma\left(\frac{1-\nu}{2}\right)\right]-\nu\ln 2-\ln 2\pi,
\end{align}
which can be obtained by using \eqref{3.5}, noting $m_0=0$. 
On the other hand, by recalling \eqref{2.25} one has
\begin{align}\lb{4.15}
     \tfrac{d}{ds}\zeta(s;T_{0,\nu}^{(0,\pi/2)})|_{s=0} =-2\ln 2\, \zeta_{H}\left(0,\tfrac{1-\nu}{2}\right)+2\tfrac{d}{dw}\zeta_{H}\left(w,\tfrac{1-\nu}{2}\right)|_{w=0},
\end{align}
which, by using \eqref{2.28}, confirms the expression obtained in \eqref{4.14}.

For the remaining $\alpha=0=\beta$ case, we refer to the result in \eqref{2.29}, which holds for all $\mu,\nu\in[0,1)$.

\section{Concluding remarks}
In this work we have performed the analytic continuation of the spectral $\z$-function associated with some self-adjoint extensions of the Schr\"odinger operator endowed with a P\"oschl--Teller potential. We have found that for particular self-adjoint extensions, the spectral $\z$-function possesses quite an exotic structure consisting of poles located not at the usual points and of countably many logarithmic branch points (with the locations of both depending on the parameters of the operator), in addition to the usual simple poles. Such interesting structure, but in a much simpler form, has been observed before in the ambit of higher-dimensional conical manifolds but this is one of the first examples (another one can be found, as we have mentioned before, for the Legendre equation in \cite{FPS24}) in which this occurs in one dimension.   
One of the main findings of this work consists in the observation that the structure of the spectral $\z$-function for some quasi-regular (and, by extension, regular) Sturm--Liouville operators heavily depends of the particular self-adjoint extension chosen. In fact, while the Friedrichs extension, for instance, produces a rather tame $\z$-function structure consisting of a single simple pole, other separated self-adjoint extensions lead to a more exotic structure consisting of countably many simple poles and logarithmic branch points (see Theorem \ref{thmstructure}).  

This is to be contrasted to the case of quasi-regular Sturm--Liouville operators that can be transformed into Schr\"odinger operators with smooth potentials on a finite interval, where the spectral $\z$-function extends to a meromorphic function with isolated simple poles regardless of the self-adjoint extension chosen. It would be very interesting to study if this exotic behavior of the spectral $\z$-function can also be found for Sturm--Liouville problems with trace-class resolvents where at least one endpoint is limit point. It would be of particular interest to analyze whether this exotic behavior could happen for a problem with two limit point endpoints. Since in this case the operator possesses only one self-adjoint extension, any exotic behavior would correspond to the problem itself rather than the choice of a particular extension as seen in the current study. We hope to report on this matter in a future work.

\appendix
\section{The Hypergeometric and P\"oschl--Teller Differential Equations} \lb{sA}

In this appendix we provide the connection between the hypergeometric differential equation
\begin{align}\lb{A.1}
\xi(1-\xi)\dfrac{d}{d\xi^2}w(\xi)+[c-(a+b+1)\xi]\dfrac{d}{d\xi}w(\xi)-abw(\xi)=0, \quad \xi \in (0,1),
\end{align}
(cf. \cite[Sec. 15.10]{DLMF}) and the Schr\"odinger differential equation with P\"oschl--Teller potential
\begin{align}\lb{A.2}
\begin{split}
\tau_{\mu,\nu}y(z,x) = - y''(z,x) +\left(\dfrac{\mu^2-(1/4)}{\sin^2 (x)}+\dfrac{\nu^2-(1/4)}{\cos^2 (x)}\right)y(z,x)=z y(z,x), \\
\mu,\nu \in [0,\infty), \; x\in(0,\pi/2),\; z\in\bbC.
\end{split}
\end{align}
To this end, we begin with the change of variable $\xi=\sin^2(x)$ in \eqref{A.2} to arrive at
\begin{align}\lb{A.3}
\begin{split}
-4\xi(1-\xi)\dfrac{d}{d\xi^2}y(z,\xi)-2(1-2\xi)\dfrac{d}{d\xi}y(z,\xi) +\left(\dfrac{\mu^2-(1/4)}{\xi}+\dfrac{\nu^2-(1/4)}{1-\xi}\right)y(z,x)=z y(z,\xi), \\
\mu,\nu \in [0,\infty), \; \xi\in(0,1),\; z\in\bbC.
\end{split}
\end{align}
(A similar analysis holds when letting $\xi=\cos^2(x)$ instead.) For the sake of notation, we will allow values of $\pm\mu$ in the following while fixing $+\nu$, with the understanding that analogous statements hold for $-\nu$. Now, introducing the change of dependent
variable
\begin{equation}\label{A.4}
y\mapsto u,\quad u(z,\xi)=\xi^{(-1\mp2\mu)/4} (1-\xi)^{(-1-2\nu)/4} y(z,\xi),\quad \xi\in(0,1),
\end{equation}
transforms \eqref{A.3} into (after dividing through to simplify)
\begin{align}
\begin{split}\label{A.5}
\xi(1-\xi)\dfrac{d}{d\xi^2}u(z,\xi)&+[1\pm\mu-([(1\pm2\mu)/2]+[(1+2\nu)/2]+1)\xi]\dfrac{d}{d\xi}u(z,\xi)\\
&-(1/4)\big[(1\pm\mu+\nu)^2-z\big]u(z,x)=0,\quad \mu,\nu \in [0,\infty), \; \xi\in(0,1),\; z\in\bbC.
\end{split}
\end{align}
Comparing \eqref{A.1} and \eqref{A.5} yields $c=1\pm\mu$ and the system of equations
\begin{align}
\begin{split}\label{A.6}
a+b&=[(1\pm2\mu)/2]+[(1+2\nu)/2],\\
ab&=(1/4)\big[(1\pm\mu+\nu)^2-z\big].
\end{split}
\end{align}
Therefore, one readily verifies that the choice
\begin{equation}
a=\dfrac{1\pm\mu+\nu+z^{1/2}}{2},\quad b=\dfrac{1\pm\mu+\nu-z^{1/2}}{2},
\end{equation}
satisfies \eqref{A.6}. Once again, the exact same results hold by replacing $\nu$ with $-\nu$ everywhere. Next, we recall the solutions of \eqref{A.1} near $\xi=0$ (cf. \cite[Eqs.~15.10.2, 15.10.3, 15.10.12]{DLMF})
\begin{align}\label{A.8}
w_{1,0}(\xi) &= \mathstrut_2F_1(a,b;c;\xi) = \sum_{n \in \bbN_0} \f{(a)_n (b)_n}{(c)_n} \f{\xi^n}{n!}, \quad 
a, b \in \bbC, \; c \in \bbC \backslash (-\bbN_0), \; \xi \in (0,1),   \no \\
w_{2,0}(\xi) &= \xi^{1-c}\mathstrut_2F_1(a-c+1,b-c+1;2-c;\xi)\\
&=\xi^{1-c}(1-\xi)^{c-a-b}\mathstrut_2F_1(1-a,1-b;2-c;\xi),\quad a, b \in \bbC, \; (c-1) \in \bbC \backslash \bbN,\; \xi \in (0,1),     \no 
\end{align}
which are linearly independent if $(a-b), (c-a-b), c \in \bbC \backslash \bbZ$. Of course, for different choices of parameters, logarithmic solutions will be required. In the special case of importance here, when $c=1$ (i.e., $\mu=0$ above) one has, from \cite[Eq. 15.5.17]{AS72},
\begin{align}\label{A.9}
w_{1,0}(\xi) &= \mathstrut_2F_1(a,b;1;\xi), \quad a, b \in \bbC, \; \xi \in (0,1),  \no \\
w_{2,0}^{\ln}(\xi) &= \mathstrut_2F_1(a,b;1;\xi) \, \ln(\xi)+\sum_{n\in\bbN}\f{(a)_n(b)_n}{(n!)^2}\xi^n\\
&\quad\, \times[\psi(a+n)-\psi(a)+\psi(b+n)-\psi(b)-2\psi(n+1)-2\gamma_E], \quad a,b \in \bbC \backslash (- \bbN_0),\; \xi \in (0,1),  \no  
\end{align}
where the superscript ``$\ln$'' indicates the presence of a logarithmic term (familiar from Frobenius theory). Once again, the previous considerations also hold with the replacement of $\nu$ with $-\nu$ everywhere.

This leads to the linearly independent solutions of \eqref{A.2} in multiple forms through \eqref{A.4}. The explicit expressions we will use are
\begin{align}
\begin{split}\label{A.10}
y_1(z,x)&=[\sin(x)]^{(1+2\mu)/2}[\cos(x)]^{(1+2\nu)/2}\\
&\quad\times\mathstrut_2F_1\big(\big(1+\mu+\nu+z^{1/2}\big)/2,\big(1+\mu+\nu-z^{1/2}\big)/2;1+\mu;\sin^2(x)\big),\quad \mu\in[0,1),
\end{split}\\
\begin{split}\notag
y_2(z,x)&=\begin{cases}
[\sin(x)]^{(1-2\mu)/2}[\cos(x)]^{(1-2\nu)/2}\\
\quad \times\mathstrut_2F_1\big(\big(1-\mu-\nu+z^{1/2}\big)/2,\big(1-\mu-\nu-z^{1/2}\big)/2;1-\mu;\sin^2(x)\big), & \mu\in(0,1),\\[2mm]
[\sin(x)]^{1/2}[\cos(x)]^{(1+2\nu)/2}\\
\ \times \bigg\{\mathstrut_2F_1\big(\big(1+\nu+z^{1/2}\big)/2,\big(1+\nu-z^{1/2}\big)/2;1;\sin^2(x)\big) \, \ln\big(\sin^2(x)\big)\\
\displaystyle \quad +\sum_{n\in\bbN}\f{\big(\big(1+\nu+z^{1/2}\big)/2\big)_n\big(\big(1+\nu-z^{1/2}\big)/2\big)_n}{(n!)^2}\big(\sin^2(x)\big)^n \\
\quad\times\big[\psi\big(n+\big(\big(1+\nu+z^{1/2}\big)/2\big)\big)-\psi\big(\big(1+\nu+z^{1/2}\big)/2\big) \\
\quad\ +\psi\big(n+\big(\big(1+\nu-z^{1/2}\big)/2\big)\big)-\psi\big(\big(1+\nu-z^{1/2}\big)/2\big)-2\psi(n+1)-2\gamma_E\big]\bigg\}, & \mu=0,
\end{cases}
\end{split}\\
&\hspace{8cm} \nu\in[0,1),\; x\in(0,\pi/2),\; z\in\bbC.
\end{align}
Applying the normalization \eqref{2.6} now yields \eqref{2.7} and \eqref{2.8}.

\subsection{Connection formulas}\label{sA1}

In this work we use connection formulas that are needed to find the behavior at $x=\pi/2$ of the solutions of \eqref{A.2} normalized at $x=0$. This reduces to the analysis of the connection formulas for the hypergeometric functions given in \eqref{A.8} and \eqref{A.9}. So we begin by introducing linearly independent solutions at $\xi=1$ of the hypergeometric differential equation \eqref{A.1} by the change of parameters
\begin{equation}
    (a, \, b, \, c, \, \xi) \rightarrow (a, \, b, \, a+b-c+1, \, 1 - \xi),
\end{equation}
or, explicitly,
\begin{align}
    w_{1,1}(\xi) &= \mathstrut_2F_1(a,b;a+b-c+1;1-\xi), 
    \\ 
    w_{2,1}(\xi) &= (1-\xi)^{c-a-b}\mathstrut_2F_1(c-a,c-b;c-a-b+1;1-\xi), \\
        w_{2,1}^{\ln}(\xi) &= \mathstrut_2F_1(a,b;1;1-\xi)\ln(1-\xi)+\sum_{n\in\bbN}\f{(a)_n(b)_n}{(n!)^2}(1-\xi)^n  \\
        &\quad\, \times[\psi(a+n)-\psi(a)+\psi(b+n)-\psi(b)-2\psi(n+1)-2\gamma_E]. \no
\end{align}
The typical connection formulas are well-known (see \eqref{2.12} and \eqref{2.13}), however, the connection formulas for the logarithmic cases do not appear to be included in the standard references.
For the logarithmic case $c=1$, we instead refer to \cite{GLPS23}. In particular, \cite[(B.16)]{GLPS23} gives the connection formula for $\mu=0\neq\nu$,
\begin{align}\label{A.16}
\begin{split}
    w_{2,0}^{\ln}(\xi) &=- [\psi(1-a)+\psi(1-b)+2\g_{E}] \frac{\G(1-a-b)}{\G(1-a)\G(1-b)}w_{1,1}(\xi)
    \\
    & \quad \, - [\psi(a)+\psi(b)+2\g_E] \frac{\G(a+b-1)}{\G(a)\G(b)}w_{2,1}(\xi) \\
    &\underset{x\uparrow1}{=} -\; [\psi(1-a)+\psi(1-b)+2\g_{E}] \frac{\G(1-a-b)}{\G(1-a)\G(1-b)}
    \\
    & \quad \, - [\psi(a)+\psi(b)+2\g_E] \frac{\G(a+b-1)}{\G(a)\G(b)}(1-\xi)^{c-a-b},
\end{split}
\end{align}
whereas the exceptional case $\mu=\nu=0$ is given in \cite[(B.23)]{GLPS23} as
\begin{align}\label{A.17}
\begin{split}
    w_{2,0}^{\ln}(\xi) &=  \pi^{-1} \sin(\pi a) \big\{\big[[\psi(a)+\psi(b)+2\g_E]^2-\pi^2 [\sin(\pi a)]^{-2}
\big] w_{1,1}(\xi)
    \\
    &\quad + [\psi(a)+\psi(b)+2\g_E] w_{2,1}^{\ln}(\xi)\big\}\\
&\underset{x\uparrow1}{=}  \pi^{-1} \sin(\pi a) \big\{\big[[\psi(a)+\psi(b)+2\g_E]^2-\pi^2 [\sin(\pi a)]^{-2}
\big]
    \\
    &\quad + [\psi(a)+\psi(b)+2\g_E] \ln(1-\xi)\big\}.
\end{split}
\end{align}
Hence, letting $\xi=\sin^2(x)$ and choosing $a,b$ in \eqref{A.16} and \eqref{A.17} as in \eqref{A.10} with $\mu=0$, and multiplying by the prefactor 
\begin{equation}
-2^{-1}[\sin(x)]^{1/2}[\cos(x)]^{(1-2\nu)/2},
\end{equation}
now allows one to apply the boundary values \eqref{2.4} and \eqref{2.5} to the logarithmic solution $\theta_{0,\nu}$.

\section{Asymptotic expansion of the characteristic functions} \lb{sB}

In this appendix, we outline the large-$z$ asymptotic expansions of the characteristic functions considered in Section \ref{s3} which are used in the process of analytic continuation of the associated spectral $\z$-functions. 
We focus, first, on the large-$z$ asymptotic expansion, with $\Im(z^{1/2})>0$ of the characteristic function $F_{0,\nu}^{(\a,\b)}(z)$ and its logarithm which is needed in the integral representation \eqref{19a}. To this end, it is convenient to consider the large-$z$ expansion of expressions that serve as the building blocks of $F_{0,\nu}^{(\a,\b)}(z)$. 

\subsection{Gamma function expansions}\lb{SubB1}
First, we outline the large-$z$ asymptotic expansion for $\Im(z^{1/2})>0$ 
and $y\in\R$, of the expression
\begin{equation}
    \left[\Gamma(y+z^{1/2}/2)\Gamma(y-z^{1/2}/2)\right]^{-1}.
\end{equation}
By using \cite[Eq. 5.5.3]{DLMF} one obtains
\begin{equation}
    \left[\Gamma(y+z^{1/2}/2)\Gamma(y-z^{1/2}/2)\right]^{-1}=\frac{1}{\pi}\sin\left[\pi(z^{1/2}/2+1-y)\right]
    \frac{\Gamma(1-y+z^{1/2}/2)}{\Gamma(y+z^{1/2}/2)}.
\end{equation}
By noticing that for $\Im(z^{1/2})>0$
\begin{equation}\lb{B3}
 \sin\left[\pi(z^{1/2}/2+1-y)\right]=\frac{1}{2}e^{-i\pi z^{1/2}/2}e^{i\pi(y-1/2)}\left[1+O\left(e^{i\pi z^{1/2}}\right)\right],  
\end{equation}
and, according to \cite[Eq. 5.11.13]{DLMF}, 
\begin{equation}\lb{B4}
  \frac{\Gamma(1-y+z^{1/2}/2)}{\Gamma(y+z^{1/2}/2)}=z^{1/2-y}2^{2y-1}\left[\sum_{k=0}^{N}\frac{2^{k}G_{k}(1-y,y)}{z^{k/2}}+O(z^{-(N+1)/2})\right],   
\end{equation}
where the functions $G_{k}$ can be expressed in terms of generalized Bernoulli polynomials $B_{n}^{(l)}(x)$, given in \cite[Sec. 24.16(i)]{DLMF}, as follows \cite[Eq. 5.11.17]{DLMF}
\begin{equation}
    G_{k}(x,y)=\binom{x-y}{k} B_{k}^{(x-y+1)}(x).
\end{equation}
The expansions \eqref{B3} and \eqref{B4} allow us to conclude that for $z\to\infty$ with $\Im(z^{1/2})>0$
\begin{align}\lb{B6}
\left[\Gamma(y+z^{1/2}/2)\Gamma(y-z^{1/2}/2)\right]^{-1}&=\frac{z^{1/2-y}2^{2y-1}}{2\pi}e^{-i\pi z^{1/2}/2}e^{i\pi(y-1/2)}\nonumber\\
&\quad\times\left[\sum_{k=0}^{N}\frac{2^{k}G_{k}(1-y,y)}{z^{k/2}}+O(z^{-(N+1)/2})\right].
\end{align}
By using the reflection formula for the generalized Bernoulli polynomials (see, e.g., \cite{ST88}), we obtain
\begin{equation}
 G_{k}(1-y,y)=\binom{1-2y}{k} B_{k}^{2(1-y)}(1-y)=(-1)^{k}\binom{1-2y}{k} B_{k}^{2(1-y)}(1-y)=(-1)^{k}G_{k}(1-y,y),
\end{equation}
which implies that the polynomials $G_{k}$ with odd index vanish identically and \eqref{B6} simplifies to 
\begin{align}\lb{B8}
\left[\Gamma(y+z^{1/2}/2)\Gamma(y-z^{1/2}/2)\right]^{-1}&=\frac{z^{1/2-y}2^{2y-1}}{2\pi}e^{-i\pi z^{1/2}/2}e^{i\pi(y-1/2)}\nonumber\\
&\quad\times\left[1+\sum_{k=1}^{N}\frac{2^{2k}G_{2k}(1-y,y)}{z^{k}}+O(z^{-N-1})\right].
\end{align}

\subsection{Digamma function expansions}\lb{SubB2}
We next analyze the following sum of digamma functions,
\begin{equation}
    \psi(y+z^{1/2}/2)+\psi(y-z^{1/2}/2),
\end{equation}
with $y$ playing the same role as above. By using the relation \cite[Eq. 5.5.4]{DLMF} one obtains 
\begin{equation}\lb{B10}
\psi(y+z^{1/2}/2)+\psi(y-z^{1/2}/2)=\psi(y+z^{1/2}/2)
+\psi(1-y+z^{1/2}/2)+\pi \cot\left[\pi(z^{1/2}/2+1-y)\right].
\end{equation}
The large-$z$ asymptotic expansion of the sum of the digamma functions on the right-hand side of \eqref{B10}, with $\Im(z^{1/2})>0$, can be obtained by applying \cite[Eq. 5.5.2]{DLMF} and the expansion \cite[Eq. 5.11.2]{DLMF},
\begin{align}
\psi(y+z^{1/2}/2)+\psi(1-y+z^{1/2}/2)&=2\ln(z^{1/2})-2\ln 2+\ln(1+2y/z^{1/2})+\ln(1-2y/z^{1/2})\nonumber\\
&\quad+z^{-1/2}(1-2y/z^{1/2})^{-1}-z^{-1/2}(1+2y/z^{1/2})^{-1}\\
&\quad-\sum_{k=1}^{N}\frac{B_{2k}}{2k}\left[(z^{1/2}/2+y)^{-2k}+(z^{1/2}/2-y)^{-2k}\right]+O(z^{-N-1}),\nonumber
\end{align}
where $B_k$ denote the Bernoulli numbers \cite[Sec. 24.2(i)]{DLMF}. Since 
\begin{align}
    \ln(1+2y/z^{1/2})+\ln(1-2y/z^{1/2})+z^{-1/2}(1-2y/z^{1/2})^{-1}-z^{-1/2}(1+2y/z^{1/2})^{-1}\nonumber\\
    =\sum_{k=1}^{N}\left[2(2y)^{2k-1}-\frac{(2y)^{2k}}{k}\right]z^{-k}+O(z^{-N-1}),
\end{align}
and
\begin{align}
\sum_{k=1}^{N}\frac{B_{2k}}{2k}\left[(z^{1/2}/2+y)^{-2k}+(z^{1/2}/2-y)^{-2k}\right]=\sum_{m=1}^{M}C_{m}(y)z^{-m}+O(z^{-M-1}),    
\end{align}
with
\begin{align}
    C_{m}(y)=\sum_{j=0}^{m-1}\binom{2m-1}{2j}\frac{2^{2m}y^{2j}}{m-j}B_{2(m-j)},
\end{align}
and by using the expansion
\begin{equation}
 \cot\left[\pi(z^{1/2}/2+1-y)\right]=-i+O(e^{i\pi z^{1/2}}),\quad \Im(z^{1/2})>0,  
 \end{equation}
we can write 
\begin{align}\lb{B16}
    \psi(y+z^{1/2}/2)+\psi(y-z^{1/2}/2)&=2\ln(z^{1/2})-2\ln 2-i\pi+\sum_{k=1}^{N} E_{k}(y)z^{-k}+O(z^{-N-1}),
\end{align}
where we have introduced the coefficients
\begin{align}\label{B17}
    E_{k}(y)=2(2y)^{2k-1}-\frac{(2y)^{2k}}{k}-C_k(y).
\end{align}

\subsection{Characteristic function expansion when \texorpdfstring{$\mu=0$}{mu} and \texorpdfstring{$\nu\in(0,1)$}{nu}}\lb{SubB3}
By using the expansions \eqref{B8} and \eqref{B16} we obtain the one for the following expression, which represents the main component of $F_{0,\nu}^{(\a,\b)}(z)$,
\begin{align}\lb{B19}
&\left[-\cos(\a)+\frac{\sin(\a)}{2}\big[2\gamma_E  +\psi\big(y+ z^{1/2}/2\big)+\psi\big(y - z^{1/2}/2\big)\big]\right]\left[\Gamma(y+z^{1/2}/2)\Gamma(y-z^{1/2}/2)\right]^{-1}\nonumber\\
&\quad=\frac{z^{1/2-y}2^{2y-1}}{2\pi}e^{-i\pi z^{1/2}/2}e^{i\pi(y-1/2)}[-\cos(\a)+\sin(\a)(\gamma_E+\ln(z^{1/2})-\ln 2-i\pi/2)]\nonumber\\
&\qquad\times\left[\sum_{k=0}^{N}\mathcal{P}_{k}(y,z)z^{-k}+O(z^{-N-1})\right],   
\end{align}
where
\begin{align}
\mathcal{P}_{0}(y,z)=1,\quad\mathcal{P}_{k}(y,z)=\sum_{j=0}^{k}4^{k-j}\mathcal{E}_{j}(y,z)G_{2(k-j)}(1-y,y),\quad k\geq 1,
\end{align}
and
\begin{align}
\mathcal{E}_{0}(y,z)=1\quad  \mathcal{E}_{j}(y,z)=\frac{\sin(\a) }{2[-\cos(\a)+\sin(\a)(\gamma_E+\ln(z^{1/2})-\ln 2-i\pi/2)]} E_{j}\left(y\right),\quad j\geq 1,
\end{align}
with $E_j(y)$ given in \eqref{B17}.
The first few functions $\mathcal{P}_{k}(y,z)$ have the explicit expressions
\begin{align}\lb{B21}
  \mathcal{P}_{1}(y,z)&=\frac{2}{3} y \left(2 y^2-3 y+1\right)+\frac{\sin(\a)(-12 y^2+12 y+2)}{6[-\cos(\a)+\sin(\a)(\gamma_E+\ln(z^{1/2})-\ln 2-i\pi/2)]},\nonumber\\
  \mathcal{P}_{2}(y,z)&=\frac{2}{45} y \left(20 y^5-24 y^4-25 y^3+30 y^2+5 y-6\right)\nonumber\\
  &\quad-\frac{2\sin(\a) \left(60 y^5-60 y^4-70 y^3-105 y^2-5 y+3\right)}{45[-\cos(\a)+\sin(\a)(\gamma_E+\ln(z^{1/2})-\ln 2-i\pi/2)]},\\
  \mathcal{P}_{3}(y,z)&=\frac{8}{63} (y-1) \left(35 y^2-91 y+60\right) \binom{1-2 y}{6}\nonumber\\
  &\quad -\frac{2 \sin(\a)\left(840 y^8+672 y^7-3962 y^6-6342 y^5-12215 y^4-1848 y^3+2611 y^2+84 y-120\right)}{945 [-\cos(\a)+\sin(\a)(\gamma_E+\ln(z^{1/2})-\ln 2-i\pi/2)]}.\no
\end{align}
We can proceed, now, with the evaluation of the large-$z$ asymptotic expansion of the characteristic function $F_{0,\nu}^{(\a,\b)}(z)$, given in \eqref{3.2}, and its logarithm.
By using \eqref{B19} we have, for $\Im(z^{1/2})>0$,
\begin{align}\lb{3.11}
F_{0,\nu}^{(\a,\b)}(z)&=-e^{-i\pi z^{1/2}/2}\frac{z^{\nu/2}}{\pi}[-\cos(\a)+\sin(\a)(\gamma_E+\ln(z^{1/2})-\ln 2-i\pi/2)]\nonumber\\
&\quad\times\Bigg[\Gamma(-\nu)2^{-\nu-1}e^{-i\pi\nu/2}\sin(\b)\Bigg(1+\sum_{k=1}^{N}\mathcal{P}_{k}\left(\frac{1-\nu}{2},z\right)z^{-k}\Bigg)\nonumber\\
&\quad+\Gamma(1+\nu)2^{\nu}e^{i\pi\nu/2}\cos(\b)z^{-\nu}\Bigg(1+\sum_{k=1}^{N}\mathcal{P}_{k}\left(\frac{1+\nu}{2},z\right)z^{-k}\Bigg)+O(z^{-N-1})\Bigg].
\end{align} 

At this point, we need to distinguish between two cases, namely, $\b=0$ and $\b\neq 0$.
\subsubsection{Case $\b\neq 0$}
When $\b\neq 0$ we can write
\begin{align}\lb{3.9}
\ln[z^{-m_0}F_{0,\nu}^{(\a,\b)}(z)]&= -\frac{i\pi z^{1/2}}{2}+\ln\left[-\frac{\Gamma(-\nu)\sin(\b)e^{-i\pi\nu/2}}{2^{\nu+1}\pi}\right] +\left(\frac{\nu}{2}-m_0\right) \ln z\nonumber\\
&\quad+\ln[-\cos(\a)+\sin(\a)(\gamma_E+\ln(z^{1/2})-\ln 2-i\pi/2)]\nonumber\\
&\quad+\ln\Bigg[\sum_{k=0}^{N}\mathcal{P}_{k}\left(\frac{1-\nu}{2},z\right)z^{-k}+\sum_{k=0}^{N}\bar{\mathcal{P}}_{k}\left(\frac{1+\nu}{2},z\right)z^{-k-\nu}+O(z^{-N-1})\Bigg],
\end{align}
where we have defined
\begin{align}\lb{B24}
\bar{\mathcal{P}}_{k}\left(\frac{1+\nu}{2},z\right)=\frac{2^{2\nu-1}\Gamma(1+\nu)}{\Gamma(-\nu)}\cot(\b)e^{i\pi\nu} \mathcal{P}_{k}\left(\frac{1+\nu}{2},z\right).   
\end{align}
At this point, we extract the first sum in the last logarithm of \eqref{3.9} to get 
\begin{align}\lb{B25}
\ln[z^{-m_0}F_{0,\nu}^{(\a,\b)}(z)]&= -\frac{i\pi z^{1/2}}{2}+\ln\left[-\frac{\Gamma(-\nu)\sin(\b)e^{-i\pi\nu/2}}{2^{\nu+1}\pi}\right] +\left(\frac{\nu}{2}-m_0\right) \ln z\nonumber\\
&\quad+\ln[-\cos(\a)+\sin(\a)(\gamma_E+\ln(z^{1/2})-\ln 2-i\pi/2)]\nonumber\\
&\quad+\ln\Bigg[\sum_{k=0}^{N}\mathcal{P}_{k}\left(\frac{1-\nu}{2},z\right)z^{-k}+O(z^{-N-1})\Bigg]\nonumber\\
&\quad+\ln\Bigg[1+\sum_{k=0}^{N}\Omega_{k}\left(\nu,z\right)z^{-k-\nu}+O(z^{-N-\nu-1})\Bigg],    
\end{align}
where one finds that 
\begin{align}\lb{B26}
    \Omega_{0}(\nu)&=\frac{2^{2\nu-1}\Gamma(1+\nu)}{\Gamma(-\nu)}\cot(\b)e^{i\pi\nu}, \nonumber\\
    \Omega_{k}(\nu,z)&=\bar{\mathcal{P}}_{k}\left(\frac{1+\nu}{2},z\right)-\sum_{l=0}^{k-1}\mathcal{P}_{k-l}\left(\frac{1-\nu}{2},z\right)\Omega_{l}(\nu,z),\quad k\geq 1.
\end{align}
By employing the fact
\begin{align}\lb{4.7}
    \ln\left(1+\sum_{m=1}^\infty c_m x^m\right) = \sum_{m=1}^\infty d_m x^m, \quad 0\leq |x|\text{ sufficiently small},
\end{align}
where
\begin{equation}\lb{4.9}
d_1=c_1, \quad d_j=c_j-\sum_{\ell=1}^{j-1} (\ell /j) c_{j-\ell}d_{\ell},\quad j \in \bbN, \; j\geq 2,
\end{equation}
we obtain
\begin{align}\lb{B29}
\ln\Bigg[\sum_{k=0}^{N}\mathcal{P}_{k}\left(\frac{1-\nu}{2},z\right)z^{-k}+O(z^{-N-1})\Bigg]=\sum_{m=1}^{N}\mathcal{R}_{m}(\nu,z)z^{-m}+O(z^{-N-1}),   
\end{align}
with the recurrence relation, for $m\geq 1$,
\begin{align}\lb{B30}
  \mathcal{R}_{1}(\nu,z)=\mathcal{P}_{1}\left(\frac{1-\nu}{2},z\right),\quad\mathcal{R}_{m}(\nu,z)=\mathcal{P}_{m}\left(\frac{1-\nu}{2},z\right)-\sum_{l=1}^{m-1} (l/m) \mathcal{P}_{m-l}\left(\frac{1-\nu}{2},z\right)\mathcal{R}_{l}(\nu,z).
\end{align}

\begin{remark}\label{RemB}
The large-$z$ expansion of the last logarithm in \eqref{B25} is, for arbitrary $\nu\in(0,1)$, non-trivial. In fact, one could proceed by expanding this term in the same way as it was done in \eqref{B29} by regarding the product $\Omega_k(\nu,z)z^{-\nu}$ as the $z$-dependent coefficient of $z^{-k}$. However, to find an asymptotic expression similar to the the one on the right-hand side of \eqref{B29}, one must be able to order terms of the form $z^{-k-j\nu}$, $k=0,\dots, N$, and $j=1,\dots,k$. Without giving the value for $\nu\in(0,1)$, it becomes impossible to order these terms in all generality. Let us point out, though, that for a given fixed value of $\nu\in(0,1)$, this process can be followed to the required asymptotic order.
\hfill$\diamond$
\end{remark}

An explicit expression for the asymptotic expansion of the last logarithm in \eqref{B25}  can be obtained, by following an argument similar to the one that provides \eqref{B29}, if we restrict our analysis to rational values of $\nu\in(0,1)$.
By setting $\nu=p/q$ with $p,q\in\N$, $0< p\leq q-1$, and $p,q$ relatively prime, and by introducing the set, which represents a subset of the congruence class of $p\,(\textrm{mod}\, q)$,
\begin{align}
    [p]_q=\{j\in\N\, |\, j=kq+p\; \textrm{for}\; k\in\N_{0}\}
\end{align}
one finds
\begin{align}\lb{B32}
\ln\Bigg[1+\sum_{k=0}^{N}\Omega_{k}\left(p/q,z\right)z^{-k-\frac{p}{q}}+O(z^{-N-\frac{p}{q}-1})\Bigg]=\sum_{m=0}^{N}\mathcal{S}_{m}(p/q,z)z^{-\frac{m+p}{q}}+O(z^{-\frac{N+1+p}{q}}),  
\end{align}
where for $m\geq 0$
\begin{align}\lb{B33}
\mathcal{S}_{m}(p/q,z)=\Omega_{m/q}\left(p/q,z\right)-\sum_{i=p}^{m}\frac{i}{m+p}\Omega_{\frac{m-i}{q}}\left(p/q,z\right)\mathcal{S}_{i-p}(p/q,z),    
\end{align}
with the understanding that when $m<p$ the sum vanishes and
\begin{align}
\Omega_{m/q}\left(p/q,z\right)= \begin{cases}0,&\textrm{if}\; m\notin[p]_q ,\\
\Omega_{m/q}\left(p/q,z\right),& \textrm{if}\; m\in[p]_q .
\end{cases}  
\end{align}

We can, therefore, rewrite the large-$z$ asymptotic expansion \eqref{B25} as
\begin{align}\lb{B35}
\ln[z^{-m_0}F_{0,p/q}^{(\a,\b)}(z)]&= -\frac{i\pi z^{1/2}}{2}+\ln\left[-\frac{\Gamma(-p/q)\sin(\b)e^{-i\pi p/2q}}{2^{p/q+1}\pi}\right] +\left(\frac{p}{2q}-m_0\right) \ln z\nonumber\\
&\quad+\ln[-\cos(\a)+\sin(\a)(\gamma_E+\ln(z^{1/2})-\ln 2-i\pi/2)]+\sum_{m=1}^{N}\mathcal{R}_{m}(p/q,z)z^{-m}\nonumber\\
&\quad+\sum_{m=0}^{q(N+1)-p-1}\mathcal{S}_{m}(p/q,z)z^{-\frac{m+p}{q}}+O(z^{-N-1}).    
\end{align}
We would now like to point out two important aspects of this expansion. First, as will be evident below, the portion of \eqref{B35} involving $\mathcal{R}_m$ vanishes identically when $\alpha=0$ (and only then). Second, from \eqref{B26} and \eqref{B33}, the expansion involving $\mathcal{S}_m$ vanishes identically when $\beta=\pi/2$.

By using the recurrence relation \eqref{B30} and the expressions \eqref{B21} and \eqref{B24} an induction argument shows that the functions $\mathcal{R}_{m}(\nu,z)$ have the general form
\begin{align}\label{B.36}
 \mathcal{R}_{m}(\nu,z)=\sum_{l=0}^{m}\frac{p_{m,l}(\nu)(\sin(\a))^{l}}{[-\cos(\a)+\sin(\a)(\gamma_E+\ln(z^{1/2})-\ln 2-i\pi/2)]^l},\, k\geq 1,   
\end{align}
with $p_{k,l}(\nu)$ being polynomials in the variable $\nu$. 

\begin{remark}\label{RemB2}
Notice that \eqref{B.36} equals $p_{m,0}(\nu)$ when $\alpha=0$. Moreover, if $\alpha\neq0$, $\mathcal{R}_{m}(\nu,z)$ always includes at least the $m$th term on the right-hand side of \eqref{B.36}. This can be seen by first noting that this term corresponds to the $(m-1)$st term of \eqref{B30}. Utilizing the explicit form of $\mathcal{R}_1$ yields that $p_{m,m}(\nu)=(-1)^{m-1}(5-3\nu^2)^m\neq0$ for $\nu\in(0,1)$. One infers from this observation that at least the $m$th power of the denominator of \eqref{B.36} will always be present. This is important when discussing the existence of branch points in the structure of the spectral $\zeta$-function.

In addition, the analytic continuation \eqref{3.17}, in the limit $\a\downarrow 0$, shows that $\zeta(-m;T_{0,\nu}^{(0,0)})=-m p_{m,0}(-\nu)$ for $m \in\N$. On the other hand, according to \eqref{2.24}, the spectral $\z$-function associated with the Friederichs extension is $\zeta(s;T_{0,\nu}^{(0,0)})=2^{-2s}\zeta_{H}\left(2s,(1+\nu)/2\right)$. These observations lead to the following relation in terms of Bernoulli polynomials
\begin{align}\lb{B37}
  p_{m,0}(-\nu)=\frac{2^{2m}}{m(2m+1)}B_{2m+1}\left(\frac{1+\nu}{2}\right),   
\end{align}
which can be obtained by using \cite[Eq. 25.11.14]{DLMF}. N\"{o}rlund showed \cite[p. 22]{No54} (see also \cite[Sec. 24.12(i)]{DLMF}) that $0$, $1/2$, and $1$ are the only real roots of the odd Bernoulli polynomial $B_{2n+1}(x)$ for $n\in\N$ in the interval $[0,1]$. This implies that both $p_{m,0}(-\nu)$ in \eqref{B37} and $p_{m,0}(\nu)$ are never vanishing for $\nu\in(0,1)$.   
\hfill$\diamond$
\end{remark}

The first few functions $\mathcal{R}_{m}(\nu,z)$, which we can use to extract the polynomials $p_{m,l}(\nu)$ with $1\leq m\leq 3$ and $0\leq l\leq m$, are
\begin{align}
\mathcal{R}_{1}(\nu,z)&=-\frac{1}{6} \nu  \left(\nu ^2-1\right)+\frac{\sin(\a)(5-3 \nu ^2)}{6 [-\cos(\a)+\sin(\a)(\gamma_E+\ln(z^{1/2})-\ln 2-i\pi/2)]}, \nonumber\\
\mathcal{R}_{2}(\nu,z)&=-\frac{1}{60} \nu \left(\nu ^2-1\right) \left(3 \nu ^2-7\right)+\frac{\sin(\a)(-15 \nu ^4+150 \nu ^2-240 \nu +97)}{60 [-\cos(\a)+\sin(\a)(\gamma_E+\ln(z^{1/2})-\ln 2-i\pi/2)]}\nonumber\\
&\quad\,-\frac{\sin^{2}(\a)\left(3 \nu ^2-5\right)^2}{72 [-\cos(\a)+\sin(\a)(\gamma_E+\ln(z^{1/2})-\ln 2-i\pi/2)]^2},\nonumber\\
\mathcal{R}_{3}(\nu,z)&=-\frac{1}{126} \nu  \left(\nu ^2-1\right)\left(3 \nu ^4-18 \nu ^2+31\right)\nonumber\\
&\quad\,+\frac{\sin(\a)(-21 \nu ^6+525 \nu ^4-1680 \nu ^3+2037 \nu ^2-1008 \nu +179)}{126  [-\cos(\a)+\sin(\a)(\gamma_E+\ln(z^{1/2})-\ln 2-i\pi/2)]}\nonumber\\
&\quad\,-\frac{\sin^{2}(\a)\left(15 \nu ^4-150 \nu ^2+240 \nu -97\right) \left(3 \nu ^2-5\right)}{360[-\cos(\a)+\sin(\a)(\gamma_E+\ln(z^{1/2})-\ln 2-i\pi/2)]^2}\nonumber\\
&\quad\,-\frac{\sin^{3}(\a)\left(3 \nu ^2-5\right)^3}{648 [-\cos(\a)+\sin(\a)(\gamma_E+\ln(z^{1/2})-\ln 2-i\pi/2)]^3}.
\end{align}
From the recurrence relation \eqref{B33} one realizes that the explicit expression of the functions $\mathcal{S}_{m}(p/q,z)$ depends on the particular values of $p$ and $q$. Nevertheless, we can make some statements regarding their general form.

In fact, by recalling the $x\to 0$ expansion of $\ln(1+x)$, one obtains, from \eqref{B32}, the relation
\begin{align}\lb{B39}
&\ln\Bigg[1+\sum_{k=0}^{N}\Omega_{k}\left(p/q,z\right)z^{-k-\frac{p}{q}}+O(z^{-N-\frac{p}{q}-1})\Bigg]\nonumber\\
&=\sum_{n=1}^{M}\frac{(-1)^{n+1}}{n}z^{-\frac{np}{q}}\left[\left(\sum_{k=0}^{N}\Omega_{k}\left(p/q,z\right)z^{-k}\right)^{n}\right]+O(z^{-(M+1)(p/q-N)})\nonumber\\
&=\sum_{n=1}^{M}\frac{(-1)^{n+1}}{n}\left[\sum_{k=0}^{nN}\left(\sum_{\underset{i_1+i_2+\ldots+i_n=k}{0\leq i_1,i_2,\ldots,i_n\leq k}}\Omega_{i_1}\left(p/q,z\right)\ldots\Omega_{i_n}\left(p/q,z\right)\right)\right]z^{-\frac{np}{q}-k}+O(z^{-(M+1)(p/q-N)}),    
\end{align}
which we can utilize to deduce the general form of the functions $\mathcal{S}_{m}(p/q,z)$. 
From \eqref{B26} and the relation \eqref{B19} and \eqref{B24} one finds that 
\begin{align}\lb{B40}
   \sum_{\underset{i_1+i_2+\ldots+i_n=k}{0\leq i_1,i_2,\ldots,i_n\leq k}}\Omega_{i_1}\left(p/q,z\right)\ldots\Omega_{i_n}\left(p/q,z\right) =\sum_{j=0}^{k}\frac{f_{n,j}(p/q,\beta)(\sin(\a))^{j}}{[-\cos(\a)+\sin(\a)(\gamma_E+\ln(z^{1/2})-\ln 2-i\pi/2)]^j},
\end{align}
with $f_{n,j}(\nu,\beta)$ functions that are independent of $z$. It is convenient, for the discussion that follows, to define the upper limit of the sum in \eqref{B40} to be the {\it order} of that sum.  
From \eqref{B39} one can, then, deduce that the coefficient of the term of the expansion $z^{-np/q-k}$, for $n\geq 1$ and $k\geq 0$, always contains a sum of the form \eqref{B40} of order $k$. In order to understand the general form of $\mathcal{S}_{m}(p/q,z)$, defined in \eqref{B32}, we need to describe the coefficient of the power $z^{-(m+p)/q}$ in \eqref{B39}. It is clear that different combinations of the integers $n$ and $k$ in \eqref{B39} can contribute to the power $z^{-(m+p)/q}$ with the same value of $m$. In fact, the coefficient of the power $z^{-(m+p)/q}$ consists of sums of the form \eqref{B40} with integers $n$ and $k$ satisfying the relation  
\begin{align}
    \frac{np}{q}+k=\frac{m+p}{q},\quad n\in\N_{+},\quad k,m\in\N_{0}.
\end{align}  
By setting $n-1=l$, the last expression transforms into a Diophantine equation
\begin{align}\lb{B40a}
    lp+kq=m,\quad l,k\in\N_0,
\end{align}
whose solutions exist for all $m\in\N_{0}$, since $p,q$ are relatively prime. Since we are interested only in nonnegative solutions for the order $k$ and for the integer $l$, we introduce the set of \textit{acceptable} solutions of \eqref{B40a} for a fixed value of $m$
\begin{align}\lb{B42a}
    [m]_{p,q}=\{(l,k)\in\N_{0}^{2}: \;lp+kq=m\}.
\end{align}
Let $\tilde{k}$ and $\tilde{l}$ be particular solutions of \eqref{B40a}, then the set $ [m]_{p,q}$ is not empty as long as $\lceil \tilde{l}/q \rceil-\lfloor -\tilde{k}/p \rfloor-1>0$.
For instance, $[0]_{p,q}=\{(0,0)\}$ which implies that for all $p,q\in\N$  and $0<p\leq q-1$
\begin{align}\lb{B38}
 \mathcal{S}_{0}(p/q,z)= \Omega_{0}\left(p/q\right). 
\end{align}
In addition, the relation \eqref{B42a} shows that when $0<m<p$, $[m]_{p,q}=\emptyset$, which implies that $\mathcal{S}_{m}(p/q,z)=0$ for $0<m<p$. 

For $m\geq p$, the set  $[m]_{p,q}$ may or may not be empty. If it is empty, then the corresponding function $\mathcal{S}_{m}(p/q,z)=0$. If $[m]_{p,q}$ is not empty, then it might contain more than one element (since it is possible to have more than one nonnegative pair of solutions of \eqref{B40a}). Let $(l_i,k_i)\in[m]_{p,q}$ with $i\in\N$, then the functions $\mathcal{S}_{m}(p/q,z)$ are constructed from a linear combination of sums of the form \eqref{B40} of order $k_i$ with  $i\in\N$. The order of $\mathcal{S}_{m}(p/q,z)$ is, then, $k_m=\textrm{max}\{k_i\}$ with $i\in\N$, namely the {\it largest} value of $k_i$ amongst the vectors $(l_i,k_i)\in[m]_{p,q}$.

Of particular interest are the functions $\mathcal{S}_{m}(p/q,z)$ of order zero since they depend on $\Omega_{0}\left(p/q\right)$ only and are, consequently, independent of the variable $z$. The functions $\mathcal{S}_{m}(p/q,z)$ of order zero are those, according to \eqref{B40a}, for which $m=lp$ with $1\leq l\leq q-1$ (in fact, if $l=q+j$, $j\in\N_0$, then $m=qp+jp$ which introduces a term of order $p>0$). In more detail, one finds
\begin{align}\lb{B42}
 \mathcal{S}_{lp}(p/q,z)= \frac{(-1)^{l}}{l+1}[\Omega_{0}\left(p/q\right)]^{l+1},\quad 0\leq l\leq q-1. 
\end{align}
By separating the terms of order zero from those of order $k\geq 1$, which according to \eqref{B42a} can only occur for $m\geq q$, and by introducing $q_N=\textrm{min}\{q,\lceil q(N+1)/p\rceil-1\}$, we can rewrite \eqref{B35} as
\begin{align}\lb{B45a}
\ln[z^{-m_0}F_{0,p/q}^{(\a,\b)}(z)]&=\mathcal{L}^{(\a,\b)}_{asy,N}(z,0,p/q)+O(z^{-N-1}),
\end{align}
where for $\beta\in(0,\pi)$,
\begin{align}\lb{B45b}
\mathcal{L}^{(\a,\b)}_{asy,N}(z,0,p/q)&= -\frac{i\pi z^{1/2}}{2}+\ln\left[-\frac{\Gamma(-p/q)\sin(\b)e^{-i\pi p/2q}}{2^{p/q+1}\pi}\right] +\left(\frac{p}{2q}-m_0\right) \ln z\\
&\quad+\ln[-\cos(\a)+\sin(\a)(\gamma_E+\ln(z^{1/2})-\ln 2-i\pi/2)]+\sum_{m=1}^{N}\mathcal{R}_{m}(p/q,z)z^{-m}\nonumber\\
&\quad+\sum_{m=1}^{q_N}\frac{(-1)^{m-1}}{m}[\Omega_{0}\left(p/q\right)]^{m}z^{-\frac{mp}{q}}+\sum_{\underset{m\neq jp\,,0\leq j\leq q_N-1}{m=q}}^{q(N+1)-p-1}\mathcal{S}_{m}(p/q,z)z^{-\frac{m+p}{q}}.\nonumber
\end{align}

The remarks above and the expression \eqref{B40} imply that the only non-vanishing functions $\mathcal{S}_{m}(p/q,z)$ in the last sum of \eqref{B45a} are those for which $[m]_{p,q}\neq\emptyset$. These functions are of order $k_m$ and, hence, have the form 
\begin{align}\lb{B45}
 \mathcal{S}_{m}(p/q,z)=\sum_{j=0}^{k_m}\frac{g_{m,j}(p/q,\beta)(\sin(\a))^{j}}{[-\cos(\a)+\sin(\a)(\gamma_E+\ln(z^{1/2})-\ln 2-i\pi/2)]^j}, 
\end{align}
where the functions $g_{m,j}(p/q,\beta)$ can only be determined explicitly once $p$ and $q$ are given specific values.

\begin{remark}\label{RemB3}
We point out that proving that $g_{m,j}(p/q,\beta)\neq0$ is important as the $j=0$ and $j\neq0$ terms correspond to existence of poles and branch points, respectively, in \eqref{3.12}. However, unlike the case of $p_{m,j}(\nu)$ studied in Remark \ref{RemB2}, proving whether $g_{m,j}$ are nonzero does not appear generally possible. Once $p,q,\beta$ are given, one can study the structure in detail. From calculations performed for particular choices of $p$, $q$, and $\b$, we in fact conjecture that all of the $g_{m,j}(p/q,\beta)$ are nonzero for the choices of $p,\ q,$ and $\beta$ considered here. But as we cannot prove this, part of our study in Sections \ref{sub3.1.1} and \ref{sub3.1.2} must be reduced to the case when $g_{m,j}(p/q,\beta)\neq0$, which we refer to as the \textit{generic} case. This is the case we suspect to be generally true.
\hfill$\diamond$
\end{remark}

\subsubsection{Case $\b=0$}
When $\b=0$, the logarithm of the characteristic function in \eqref{3.11} reduces to 
\begin{align}\lb{B46}
\ln[z^{-m_0}F_{0,\nu}^{(\a,0)}(z)]&= -\frac{i\pi z^{1/2}}{2}+\ln\left[-\frac{\Gamma(\nu+1)e^{i\pi\nu/2}2^{\nu}}{\pi}\right] -\left(\frac{\nu}{2}+m_0\right) \ln z\nonumber\\
&\quad+\ln[-\cos(\a)+\sin(\a)(\gamma_E+\ln(z^{1/2})-\ln 2-i\pi/2)]\nonumber\\
&\quad+\ln\Bigg[1+\sum_{k=1}^{N}\mathcal{P}_{k}\left(\frac{1+\nu}{2},z\right)z^{-k}+O(z^{-N-1})\Bigg].
\end{align}
By utilizing \eqref{B29}, with the change $\nu\to-\nu$ one immediately finds
\begin{align}\lb{B48}
   \ln[z^{-m_0}F_{0,\nu}^{(\a,0)}(z)]&=\mathcal{L}^{(\a,0)}_{asy,N}(z,0,\nu)+O(z^{-N-1}),
\end{align}
where
\begin{align}\lb{B51}
 \mathcal{L}^{(\a,0)}_{asy,N}(z,0,\nu)&=  -\frac{i\pi z^{1/2}}{2}+\ln\left[-\frac{\Gamma(\nu+1)e^{i\pi\nu/2}2^{\nu}}{\pi}\right] -\left(\frac{\nu}{2}+m_0\right) \ln z\nonumber\\
&\quad+\ln[-\cos(\a)+\sin(\a)(\gamma_E+\ln(z^{1/2})-\ln 2-i\pi/2)]\nonumber\\
&\quad+\sum_{m=1}^{N}\mathcal{R}_{m}(-\nu,z)z^{-m}. 
\end{align}

\medskip


\medskip

 
\end{document}